\documentclass[
twocolumn,
superscriptaddress,
footinbib,
secnumarabic,
nobibnotes,
bibnotes,aps,
prb, longbibliography,
floatfix]{revtex4-2}
\usepackage{amsmath}
\usepackage{natbib}
\usepackage{hyperref}
\usepackage{amscd,latexsym}
\usepackage{mathrsfs}
\usepackage{graphicx}
\usepackage{epstopdf}
\usepackage{cancel}
\usepackage{amsfonts}
\usepackage{exscale}
\usepackage{dcolumn}
\usepackage{bm}
\usepackage{color}
\usepackage{lmodern}
\usepackage[T2A,T1]{fontenc}
\usepackage[latin9]{inputenc}
\usepackage{verbatim}
\usepackage{float}
\usepackage{footnote}
\usepackage{esint}
\usepackage{mathtools}
\usepackage[caption=false]{subfig}
\usepackage{centernot}

\makeatother
\DeclareMathOperator{\notmu}{\mathrel{\not\mskip-\thinmuskip \mu}}

\begin{document}

\title{Magnetic quantum oscillations of in-plane Hall conductivity and
magnetoresistance tensor in quasi-two-dimensional metals}
\date{\today }

\begin{abstract}
We develop the theory of magnetoresistance oscillations in layered
quasi-two-dimensional (quasi-2D) metals. Using the Kubo-Streda formula, we
calculate the Hall intralayer conductivity in a magnetic field perpendicular to
conducting layers. The analytical expressions for the amplitudes and
phases of magnetic quantum oscillations (MQO) and of the difference or the so-called slow
oscillations (SO) are derived as a function of several parameters: 
magnetic field strength, interlayer transfer integral, temperature, 
the Landau-level broadening and the electron mean-free time. 
We calculate the quantum oscillations of the magnetoresistance tensor, 
because the magnetoresistance rather than conductivity is usually measured. 
We also discuss the averaging of magnetoresistance oscillations 
over temperature and long-range sample inhomogeneities.
The results obtained are useful to analyze experimental data on 
magnetoresistance oscillations in various quasi-2D metals.
\end{abstract}

\author{P.D. Grigoriev}
\affiliation{L. D. Landau Institute for Theoretical Physics, Chernogolovka,
	142432, Russia} 
\affiliation{National University of Science and Technology
	<<MISiS>>, Moscow, 119049, Russia} 
\affiliation{Kotelnikov Institute of Radioengineering and Electronics of RAS, 125009 Moscow, Russia} 
\author{T.I. Mogilyuk}
\affiliation{National Research Centre <<Kurchatov Institute>>, Moscow,
	123182, Russia}
\maketitle


\section{Introduction}

Layered quasi-two-dimensional (quasi-2D) metals form a vast class of
materials, promising both for research and applications. This class includes
high-$T_{c}$ superconductors, both of cuprate and iron-based families,
organic metals, various van-der-Waals crystals, intercalated graphite,
artificial heterostructures, transition metal and rare-earth chalcogenides,
etc. A huge amount of current research is devoted to the study of
electronic structure in these materials. There are two main methods 
to measure the electronic structure in the bulk of these materials:
angle-resolved photoemission spectroscopy (ARPES) \cite{Zhang2022} and
magnetic quantum oscillations (MQO) \cite%
{Shoenberg2009Sep,Abrikosov1988Fundamentals,Ziman1972Principles}. ARPES is
more visual than MQO, but has a poorer energy and momentum resolution and
requires a high quality sample surface and sophisticated technique. The
APRES resolution is often insufficient to study fine energy splitting and
Fermi-surface (FS) reconstruction by electronic phase transitions. The MQO
provide a tool to measure the FS geometry with much higher accuracy than
ARPES and also give information about the effective mass, mean free time and
g-factor of charge carries \cite%
{Shoenberg2009Sep,Abrikosov1988Fundamentals,Ziman1972Principles}. However,
the extraction of this information from the experimental data on MQO
requires a quantitative theory of MQO. The characteristic feature of MQO is
their high sensitivity to temperature and disorder, including macroscopic
sample inhomogeneities.

The electron spectrum in layered quasi-2D metals is usually described by a
three-dimensional tight-binding dispersion 
\begin{equation}
	\epsilon _{3D}(\boldsymbol{k})=\epsilon _{\parallel }\left( \boldsymbol{k}%
	_{\parallel }\right) -2t_{z}\cos (k_{z}d),  \label{e3Dg}
\end{equation}%
where $t_{z}$ is the interlayer transfer integral of electrons and $d$ is
the interlayer distance. The in-plane electron dispersion $\epsilon
_{\parallel }\left( \boldsymbol{k}_{\parallel }\right) $ is often isotropic
and can be approximated by a parabolic one:%
\begin{equation}
	\epsilon _{\parallel }\left( \boldsymbol{k}_{\parallel }\right) =\hbar
	^{2}\left( k_{x}^{2}+k_{y}^{2}\right) /(2m_{\ast }),  \label{ePar}
\end{equation}%
where $m_{\ast }$ is the effective (cyclotron) electron mass. The FS of
layered metals related to Eq. (\ref{e3Dg}) is a corrugated cylinder. This
Fermi surface has two close extremal cross-sections areas $S_{1}$ and $S_{2}$
by the planes in $k$ space perpendicular to the magnetic field $\mathbf{
\mathit{B}}$. 
The standard Lifshitz-Kosevich (L-K)  theory in quasi-2D metals 
predicts that for each FS pocket the MQO are 
determined by the sum of oscillations with two close fundamental frequencies 
$F_{1,\,2}=cS_{1,\,2}/(2\pi |e|\hbar )$ and almost equal amplitudes 
\cite{Shoenberg2009Sep,Abrikosov1988Fundamentals,Ziman1972Principles}. 
This leads to the amplitude beats of MQO with average frequency 
$F=(F_{1}+F_{2})/2$. The beat
frequency $\Delta F\equiv F_{1}-F_{2}\approx 4t_{z}B_{z}/(\hbar \omega _{c})$
can be used to measure the interlayer transfer integral $t_{z}$ \cite%
{Shoenberg2009Sep,Kartsovnik2004Nov,Singleton2000Aug,Wosnitza2013Fermi,Kartsovnik2005Mar}%
. The phases of MQO beats in transport (across the layers) and thermodynamic quantities 
differ \cite{Grigoriev2002Jan}. This phase shift increases with magnetic
field \cite{Grigoriev2002Jan,Grigoriev2003Apr,Mogilyuk20243d2d} and reaches almost $\pi /2$
in some quasi-2D metals \cite{Balthes2000,Mogilyuk20243d2d}. 

The particular angular
dependence of the beat frequency $\Delta F$, i.e. its dependence on the
tilt angle $\theta $ of magnetic field from the normal to conducting $%
(x,\,y)$ layers, allows to evaluate the in-plane Fermi momentum $k_{F}$ \cite%
{Kartsovnik2004Nov,Singleton2000Aug,Wosnitza2013Fermi,Grigoriev2002Jan,Grigoriev2003Apr,Kartsovnik2005Mar}%
: 
\begin{equation}
	\Delta F(\theta )/\Delta F(0)=t_z(\theta )/t_z (0)=J_{0}(k_{F}d\tan (\theta
	)).  \label{beat-freq}
\end{equation}%
The beat frequency vanishes
in the so-called Yamaji angles $\theta _{\text{Yam}}$ found from the zeros
of the Bessel function in Eq. (\ref{beat-freq}). The
angular oscillations of the effective interlayer transfer integral $t_{z}(\theta )$
described by Eq. (\ref{beat-freq}) also lead to angular oscillations of
magnetoresistance (AMRO). AMRO were first discovered in a quasi-2D organic
metal $\beta $-(BEDT-TTF)$_{2}$IBr$_{2}$ in 1988 \cite{Kartsovnik1988Transverse-II}
and explained by Yamaji using geometrical arguments \cite{Yamaji1989May}.
This effect was quantitatively described next year using the Boltzmann
transport equation \cite{Yagi1990Sep}. Subsequently, AMRO were widely
analyzed in both quasi-2D and quasi-one-dimensional organic metals \cite%
{Kartsovnik2004Nov,Singleton2000Aug,Kartsovnik2005Mar,Wosnitza2013Fermi,Lebed2008The,Yagi1990Sep,Kurihara1992Mar,Moses1999Sep,Grigoriev2014Sep,Grigoriev2002Jan} 
to extract useful information about the electronic dispersion, as well as in
many other layered materials, including high-Tc superconductors \cite{Hussey2003Oct} and 
artificial heterostructures \cite{KURAGUCHI2003}. The interplay
of angular and quantum magnetoresistance oscillations is not trivial and
leads to new effects \cite{Grigoriev2014Sep,Grigoriev2017May}. These two
effects do not simply factorize in strong magnetic fields \cite%
{Grigoriev2014Sep}, as one often assumes to describe experimental results.
One of such new effects resulting from the interplay of MQO and AMRO are the
so-called false spin zeros, that may lead to incorrect estimates of the
electron g-factor from the angular dependence of MQO. This effect survives
and becomes even stronger in the presence of additional electronic states on
the Fermi surface \cite{Grigoriev2017May,Grigoriev2018Jun}, e.g., as in
multiband compounds. Note that the antiferromagnetic ground state also
strongly affects the g-factor of conducting electron and often leads to its
nullification as measured from the MQO \cite{Ramazashvili2021}.

In addition to magnetic quantum (Shubnikov) oscillations, in quasi-2D metals
there are also difference-frequency or the so-called slow oscillations (SO)
of magnetoresistance. The SO were first discovered in 1988 while studying
the organic superconductor $\beta$-(BEDT-TTF)$_2$IBr$_2$ \cite%
{Kartsovnik1988Transverse,Kartsovnik1988Transverse-II}. Similar oscillations
have also been observed in other organic conductors, e.g., $\beta$-(BEDT-TTF)%
$_2$I$_3$, $\kappa$-(BEDT-TTF)$_2$Cu$_2$(CN)$_3$ \cite{Ohmichi1998Apr}, and $%
\kappa$-(BEDT-TSF)$_2$C(CN)$_3$ \cite{Narymbetov1998Sep,Togonidze2001Jan}.
The SO were also discovered in rare-earth metal tritellurides \cite%
{Sinchenko2016Dec,Grigoriev2016Jun}. SO can also occur in a bilayer
structure despite the absence of $k_z$ dispersion, for example, in cuprate
superconductors \cite{Grigoriev2016Jun,Grigoriev2017Sep,Grigoriev2017Oct}.
The first theoretical explanation with the help of the Boltzmann equation
was given in the work \cite{Kartsovnik2002Aug}. SO arise due to the
corrugation of the Fermi surface of quasi-2D metals. Their frequency $%
F_{SO}=2\Delta F$ is equal to the doubled beat frequency. It is also
proportional to the tunnel integral $t_z$ and has the same angular
dependence given by Eq. (\ref{beat-freq}), as has been shown both
theoretically and experimentally \cite{Kartsovnik2002Aug}. It was also
shown that magnetization and seemingly other thermodynamic properties of
quasi-2D metals have no SO in the case of noninteracting electrons \cite%
{Kartsovnik2002Aug,Grigoriev2003Apr,Mogilyuk2022Aug}. These types of
oscillations are much more resistant to a temperature increase and to
macroscopic sample inhomogeneities since the SO frequency 
does not depend on the Fermi energy. Hence, the observation of SO is often
easier than of Shubnikov oscillations \cite%
{Kartsovnik2002Aug,Grigoriev2003Apr,Grigoriev2016Jun,Grigoriev2017Oct}.

From the analysis of SO, one can find the interlayer transfer integral $t_z$, giving the electron
hopping rate between the conducting layers, the mean free time $\tau_0$ of
electrons, the type of disorder in the system, etc \cite%
{Kartsovnik2002Aug,Grigoriev2003Apr,Grigoriev2017Oct}.  When the SO
originate from the $k_z$ electron dispersion the angular dependence of SO
frequency allows the evaluation of the in-plane Fermi momentum $k_F$. The SO
originating from the interlayer electron hopping resemble the magnetic
intersubband oscillations in 2D metals \cite%
{MagnetointersubbandRaikh1994,Averkiev2001} or the difference-frequency
oscillations \cite{Leeb2023} arising from several FS pockets. The main
difference between the SO and the magnetic intersubband or
difference-frequency oscillations is that both frequencies $F_{1,\, 2}
=S_{1,\,2}/(2\pi |e|\hbar )$ correspond to the same FS pocket. Hence, the SO
appear even in single-band quasi-2D metals. In addition, the effective
electron mass $m_{\ast}$ and the cyclotron frequency $\omega_c=|e|B/(m_{%
	\ast}c)$ are the same for both frequencies $F_{1,\, 2}$. Hence, the phase
and frequency of SO are not sensitive to the Fermi level, and the SO are not
damped by the smearing of Fermi level by temperature or long-range disorder,
contrary to the difference-frequency oscillations given by Eq. (29) of Ref. 
\cite{Leeb2023}. The $T$-damping of SO occurs at considerably higher
temperature and originates from the electron-phonon or electron-electron
interaction, which affects the SO amplitude via the Dingle factor. The
analytical expressions for the SO are also much simpler than for
difference-frequency oscillations corresponding to different FS pockets.

The available theory of MQO in quasi-2D metals is built only for the
diagonal magnetoconductivity, out-of-plane \cite%
{Grigoriev2002Jan,Kartsovnik2002Aug,Champel2002Nov,Grigoriev2003Apr} or
in-plane \cite{Mogilyuk2018Jul}, while the resistance is often measured
experimentally. To analyze the measured in-plane magnetoresistance, one
needs to know the Hall conductivity and invert the magnetoconductivity
tensor. Here, we calculate the quantum oscillations of Hall conductivity and
study the in-plane magnetoresistance tensor. The Hall component of
magnetoresistance is also very important, as it is often measured in layered
metals. For example, the first observation of quantum oscillations in
high-Tc cuprate superconductors in 2007 \cite{Doiron-Leyraud2007May} was on
Hall coefficient. There are many other layered materials where the experimental data
on Hall resistance oscillations require a quantitative theoretical analysis:
a bulk Bi$_2$Se$_3$ \cite{Analytis2010May,Busch2018Jan}, CaFeAsF \cite%
{Terashima2022Jun}, LaFeAsO$_{0.9}$F$_{0.1-\delta}$ \cite{Zhu2008Jul}, 
rare-earth tritellurides \cite{Higashihara2024,Sinchenko2024}, etc.

In this work we calculate the Hall conductivity and find the
intralayer magnetoresistance tensor, which contains magnetic quantum oscillation. 
In Sec. \ref{SecModel} we describe our model and approximations. In Sec. \ref{SecCond} we calculate the 
Hall conductivity component, including the quantum and differential (slow) oscillations. 
In Sec. \ref{SecR} we calculate the resistivity tensor. 
In Sec. \ref{SecDiscuss} we discuss the results, and give their summary in Sec. \ref{SecSum}. 
Appendices provide the details of our calculations.

\section{The model and basic formulas} \label{SecModel}

We consider a layered quasi-2D metal with dispersion given by Eqs. (\ref%
{e3Dg}), (\ref{ePar}) placed in a magnetic field $\mathbf{B}=(0,\,0,\,B)$
perpendicular to the conducting layers. The Landau-level (LL) quantization
of in-plane motion leads to the new electron dispersion \cite%
{Landau1991Quantum,Shoenberg2009Sep} 
\begin{equation}
	\epsilon \left(n,\,k_{z}\right) =\hbar \omega _{c}(n+1/2)-2t_z\cos (k_{z}d),
	\label{ES}
\end{equation}%
where $n$ is the LL index.

We use the Feynman diagram technique \cite{Mahan2000} and the self-consistent Born
approximation (SCBA) to find the electron self-energy function coming 
from the elastic scattering by short-range crystal defects.  
We neglect the interelectronic interaction and the
influence of phonons. Then we apply the
harmonic expansion of MQO and keep only the first- and second-order terms in
the Dingle factor to obtain the analytical expressions for conductivity
tensor. 

Previously, the MQO of diagonal interlayer $\sigma_{zz}$ and
intralayer diagonal conductivity $\sigma_{xx}$ were calculated in the same
approximations \cite{Grigoriev2003Apr,Grigoriev2014Sep,Mogilyuk2018Jul} in
quasi-2D metals.  These calculations have shown several notable differences
between the MQO of $\sigma_{zz}$ and of $\sigma_{xx}$ \cite{Mogilyuk2018Jul}%
. First, the MQO of $\sigma_{zz}$ gradually invert phase at $%
t_z\sim\hbar\omega_c/(4\pi)$ while the $\sigma_{xx}$ MQO do not. It means
that in MQO of $\sigma_{zz}$ and $\sigma_{xx}$ have the opposite phase in
the weak field and the same phase in the strong field. Second, in quasi-2D
metals there is the phase shift of beats $\sim \hbar \omega_c / t_z$ in the MQO
of $\sigma_{zz}$, which may reach $\pi /2$ \cite{Grigoriev2002Jan,Schiller2000Jul,Weiss1999Dec,Mogilyuk20243d2d}. 
In $\sigma_{xx}$ this phase
shift of beats is negligible, being smaller by the parameter $t_z/\mu\ll 1$.
Third, the amplitude of SO in $\sigma_{xx}$ depends nonmonotonically on $%
\omega_c \tau_0$ and even changes sign at $\omega_c \tau_0 \approx \sqrt{3}$ 
\cite{Mogilyuk2018Jul}. Such a phase inversion and the vanishing of
amplitude are absent in the SO of $\sigma_{zz}$. Hence, similar to MQO, the
SO of intralayer $\sigma_{xx}$ and interlayer $\sigma_{zz}$ conductivity are
in the same phase in strong field at $\omega_c \tau_0 \gg 1$ and in opposite
phase at $\omega_c \tau_0 <1 $. Does this sign change of SO amplitude also
appear in Hall conductivity $\sigma_{xy}$ and in the in-plane
magnetoresistance $\rho_{xx}$ and $\rho_{xy}$? Does the in-plane
magnetoresistance tensor have any new features? To answer these questions we
need to calculate the in-plane magnetoresistance tensor in quasi-2D metals.

\subsection{Kubo-Streda formula in quasi-2D metals}

To find the Hall conductivity of a quasi-2D metals we examine the derivation
of the Kubo-Streda formula \cite{Smrcka1977Jun} obtained for the
conductivity tensor in 2D metals (for a detailed derivation see pp. 33-42 of
Ref. \cite{Chadova2017Dec})%
\begin{equation}
	\sigma _{xy}=-\sigma _{yx}=\int_{-\infty }^{+\infty }d\varepsilon \frac{%
		dn_{F}\left( \varepsilon \right) }{d\varepsilon }\sigma _{xy}\left(
	\varepsilon \right) \equiv \sigma _{xy}^{I}+\sigma _{xy}^{II},  \label{s}
\end{equation}%
where $n_{F}\left( \varepsilon \right) =\left( 1+\exp \left[ \left(
\varepsilon -\mu \right) /T\right] \right) ^{-1}$ is the Fermi distribution
function, $\varepsilon $ is the electron energy, $\mu $ is the chemical
potential, $T$ is temperature, 
\begin{equation}
	\sigma _{\mu \nu }^{I}\left( \varepsilon \right) =\frac{-\hbar }{4\pi }%
	Tr\left\langle \hat{\jmath}_{\mu }\left( G^{A}-G^{R}\right) \hat{\jmath}%
	_{\nu }G^{R}-\hat{\jmath}_{\mu }G^{A}\hat{\jmath}_{\nu }\left(
	G^{A}-G^{R}\right) \right\rangle  \label{sI}
\end{equation}%
and 
\begin{equation}
	\sigma _{\mu \nu }^{II}\left( \varepsilon \right) =\frac{\left\vert
		e\right\vert }{4\pi i}Tr\left\langle \left( G^{A}-G^{R}\right) \left( \hat{r}%
	_{\mu }\hat{\jmath}_{\nu }-\hat{\jmath}_{\mu }\hat{r}_{\nu }\right)
	\right\rangle ,  \label{sII}
\end{equation}%
where $e=-|e|<0$ is electron charge, $G^{A}$ and $G^{R}$ are the advanced
and retarded electron Green's functions, $\hat{\jmath}_{\nu }$ is the
operator of electron current in the direction $\nu$, and $\hat{r}$ is the
coordinate operator. The trace in Eqs. (\ref{s})-(\ref{sII}) is over all
states, including all bands and the reservoir states. The derivation of Eqs.
(\ref{s})-(\ref{sII}) does not make any assumptions about the electron
dispersion and is valid also for quasi-2D multiband metals.

One can simplify the last term $\sigma _{\mu \nu }^{II}$ after integrating
it by parts and taking into account the definition of electric current $%
j_{\mu }=ev_{\mu }=e\partial \epsilon \left( \mathbf{p}\right) /\partial
p_{\mu }$, the electron Hamiltonian with Peierls substitution of the vector
potential in the symmetric gauge $\mathbf{A}=\left( -y,\,x,\,0\right) B/2$,
and the commutation relations $\left[ \hat{r}_{x},\,\hat{v}_{y}\right] =%
\left[ \hat{r}_{y},\,\hat{v}_{x}\right] =0$. As a result one obtains the
expression from the work \cite{Streda1982Aug} of P. Streda 
\begin{equation}
	\sigma _{xy}^{II}=ec\left. \frac{\partial N(\varepsilon )}{\partial B}%
	\right\vert _{\varepsilon =\mu },  \label{sIIN}
\end{equation}%
where $N(\varepsilon )$ is the total number of states below the energy $%
\varepsilon $, including all electron bands or reservoir states, and $\mu $
is the chemical potential equal to the Fermi energy. Eq. (\ref{sIIN}) can
also be used for quasi-2D metals as its derivation does not depend on the
electron dispersion and on the number of energy bands (see Appendix \ref%
{App1}).

The next well-known formula \cite{Streda1982Aug} 
\begin{equation}
	\sigma _{xy}^{I}=-\omega _{c}\tau \sigma _{xx},  \label{sIN}
\end{equation}%
where the electron mean-free time $\tau $ contains the MQO in contrast to $%
\tau_0$, is often used to calculate the Hall conductivity. The derivation of
this formula (\ref{sIN}), given in Appendix \ref{App2}, assumes a free
electron model (parabolic dispersion) and the neglects the vertex
corrections. The latter are negligible if the scattering is only by
point-like impurities. As follows from the derivation in Appendix \ref{App2}%
, Eq. (\ref{sIN}) holds for several conducting bands, and the scattering
rate $1/\tau =2 |\text{Im} \Sigma | / \hbar $ in SCBA for point-like
impurities contains the sum of the density of states from all bands and
depends on the electron energy $\varepsilon $ only.

\subsection{Harmonic expansion}

To proceed further, we assume strong harmonic damping of MQO and keep only
zeroth and first MQO harmonics. In quasi-2D metals we also keep the
SO term, which is of the second order in Dingle
factor but may be stronger than the first harmonics at high temperature or
in the presence of long-range inhomogeneity of Fermi energy. The weakly
oscillating density of electronic states (DoS) in quasi-2D metals is given
by the sum of zeroth and first harmonics \cite{Champel2001Jan}, 
\begin{equation}
	\rho \approx \rho _{0}\left[ 1-2R_{D}J_{0}(\lambda )\cos \left( \alpha
	\right) \right] ,
\end{equation}%
where the nonoscillating part of the 2D DoS (per one spin) per one band is $%
\rho_{0}=m_{\ast}/(2\pi\hbar^{2}d)$. Here the generalized Dingle factor $%
R_{D}\equiv \exp \left[ -\pi /(\omega _{c}\tau )\right] =\exp (-\gamma )$
has the MQO, because $\gamma \equiv \pi /(\omega _{c}\tau )$ oscillates, $%
\alpha =\alpha \left( \varepsilon \right) \equiv 2\pi \left[ \varepsilon -%
\text{Re}\Sigma ^{R}(\varepsilon )\right] /(\hbar \omega _{c})$ also
oscillates because the electron self-energy $\Sigma ^{R}(\varepsilon )$ has
MQO, $\lambda \equiv 4\pi t_{z}/(\hbar \omega _{c})$, $J_{0}(\lambda )$ and $%
J_{1}(\lambda )$ are the Bessel functions of the zeroth and first order.

To find $\tau $, we calculate the imaginary part of electron self-energy
function, which in SCBA is proportional to the oscillating density of
states: 
\begin{gather}
	\hbar /(2\tau )=\left\vert \text{Im}\Sigma ^{R}(\varepsilon )\right\vert =-%
	\text{Im}\Sigma ^{R}(\varepsilon )\equiv \Gamma (\varepsilon )  \label{ImS1}
	\\
	\approx \Gamma _{0}\left[ 1-2R_{D}\cos \left[ \alpha (\varepsilon )\right]
	J_{0}\left( \lambda \right) \right] .
\end{gather}%
After neglecting the second harmonics but keeping the slow-oscillating term,
this gives 
\begin{equation}
	\tau \equiv \frac{\hbar }{2\Gamma }\approx \tau _{0}\left[
	1+2R_{D}J_{0}\left( \lambda \right) \cos (\alpha )+2R_{D}^{2}J_{0}^{2}\left(
	\lambda \right) \right] .  \label{tau}
\end{equation}%
It was previously shown \cite%
{Mogilyuk2018Jul,Kartsovnik2002Aug,Grigoriev2003Apr} that the combination $%
R_{D}\cos \left( \alpha \right) $ does not contain the slow oscillations in
the second order of Dingle factor. Therefore, as we are not interested in
the second harmonic of MQO, all second-order terms coming from $R_{D}\cos
\left( \alpha \right) $ can be neglected both in the DoS and in $\tau $, and
we have%
\begin{gather}
	\rho \approx \rho _{0}\left[ 1-2R_{D0}J_{0}(\lambda )\cos \left( \overline{%
		\alpha }\right) \right] =\rho _{0}\left[ 1+\widetilde{\rho }/\rho _{0}\right]
	,  \label{rho} \\
	\tau \approx \tau _{0}\left[ 1+2R_{D0}\cos \left( \overline{\alpha }\right)
	J_{0}\left( \lambda \right) +2R_{D0}^{2}J_{0}^{2}\left( \lambda \right) %
	\right]   \notag  \label{tauII} \\
	\approx\tau _{0}\left[ 1-\widetilde{\rho }/\rho _{0}+\tau _{SO}/\tau _{0}\right] ,
	\label{tau-III}
\end{gather}%
where the Dingle factor $R_{D0}\equiv \exp \left[ -\pi /(\omega _{c}\tau
_{0})\right] $  does not oscillate, $\overline{\alpha }=2\pi \varepsilon
/(\hbar \omega _{c})$, 
$\tau _{SO}=2R_{D0}^{2}J_{0}^{2}\left( \lambda \right) \tau _{0}$, and we
introduced the notation $\widetilde{\rho }$ for the oscillating part of DoS, keeping only
its first harmonic: 
\begin{equation}
	\widetilde{\rho }/\rho _{0}\approx-2R_{D0}J_{0}(\lambda )\cos \left( \overline{%
		\alpha }\right) .  \label{TildeRho}
\end{equation}

\section{Calculations of Hall conductivity} \label{SecCond}

Experimentally one often divides the Hall conductivity $\sigma _{xy}$ into
three terms, corresponding to the monotonic part, usual MQO and the slow or
differential oscillations:%
\begin{equation}
	\sigma _{xy}(\varepsilon )\approx \overline{\sigma }_{xy}(\varepsilon
	)+\sigma _{xy}^{QO}(\varepsilon )+\sigma _{xy}^{SO}(\varepsilon ).
	\label{sxy3terms}
\end{equation}%
Below we use Eqs. (\ref{s}), (\ref{sIIN}), (\ref{sIN}) and (\ref{tau-III}) to
calculate each term in Eq. (\ref{sxy3terms}).

\subsection{Calculation of $\protect\sigma _{xy}^{I}$}

We start from Eq. (\ref{sIN}) and the necessary expressions for $\sigma _{xx}
$ at temperature $T=0$ we take from Eqs. (35), (39) and (45) of Ref. \cite%
{Mogilyuk2018Jul}, rewriting them using the relation $\gamma _{0}=\pi
/\left( \omega _{c}\tau _{0}\right) $: 
\begin{equation}
	\sigma _{xx}(\varepsilon )\approx \overline{\sigma }_{xx}(\varepsilon
	)+\sigma _{xx}^{QO}(\varepsilon )+\sigma _{xx}^{SO}(\varepsilon ).
	\label{Se3}
\end{equation}%
Here the nonoscillating term per one spin component is 
\begin{equation}
	\overline{\sigma }_{xx}(\varepsilon )\approx \frac{e^{2}}{4\pi \hbar d}\frac{%
		\overline{\alpha }\gamma _{0}}{\gamma _{0}^{2}+\pi ^{2}}=\frac{\sigma _{0}}{%
		1+\left( \omega _{c}\tau _{0}\right) ^{2}},  \label{sxxDrude}
\end{equation}%
where the Drude conductivity without magnetic field is 
\begin{equation}
	\sigma _{0}=\frac{e^{2}\overline{\alpha }\omega _{c}\tau _{0}}{4\pi
		^{2}\hbar d}=\frac{e^{2}\mu \tau _{0}}{2\pi \hbar ^{2}d}=\frac{e^{2}\tau
		_{0}n_{e}}{m_{\ast }},  \label{s0}
\end{equation}%
and the electron density $n_{e}=\mu \rho _{0}$. The quantum oscillations
term in the first order in $R_{D0}J_{0}(\lambda )$ is 
\begin{gather}
	\frac{\sigma _{xx}^{QO}(\varepsilon )}{\overline{\sigma }_{xx}}\approx
	-2R_{D0}\left[ \frac{2\pi ^{2}J_{0}\left( \lambda \right) }{\gamma
		_{0}^{2}+\pi ^{2}}\cos \left( \overline{\alpha }\right) -\frac{\lambda
		J_{1}(\lambda )}{\overline{\alpha }}\sin (\overline{\alpha })\right] 
	\label{Sq} \\
	=-2R_{D0}\left[ \frac{2\left( \omega _{c}\tau _{0}\right) ^{2}J_{0}\left(
		\lambda \right) }{1+\left( \omega _{c}\tau _{0}\right) ^{2}}\cos \left( 
	\overline{\alpha }\right) -\frac{\lambda }{\overline{\alpha }}J_{1}(\lambda
	)\sin (\overline{\alpha })\right] ,  \label{sxxOsc}
\end{gather}%
and the last term in Eq. (\ref{Se3}), responsible for SO, in the second
order in $R_{D0}J_{0}(\lambda )$ is 
\begin{gather}
	\frac{\sigma _{xx}^{SO}(\varepsilon )}{\overline{\sigma }_{xx}}\approx 2\pi
	^{2}R_{D0}^{2}J_{0}^{2}\left( \lambda \right) \frac{\pi ^{2}-3\gamma _{0}^{2}%
	}{(\gamma _{0}^{2}+\pi ^{2})^{2}}  \label{Ssl} \\
	=2R_{D0}^{2}J_{0}^{2}\left( \lambda \right) \frac{\left( \omega _{c}\tau
		_{0}\right) ^{2}\left[ \left( \omega _{c}\tau _{0}\right) ^{2}-3\right] }{%
		\left[ 1+\left( \omega _{c}\tau _{0}\right) ^{2}\right] ^{2}}.
	\label{sxxSlO}
\end{gather}%
At nonzero temperature the Eq. (\ref{sxxOsc}) should be multiplied by
temperature damping factor 
\begin{equation}
	R_{T}=2\pi ^{2}k_{B}T/(\hbar \omega _{c})/\sinh [2\pi ^{2}k_{B}T/(\hbar
	\omega _{c})].  \label{RT}
\end{equation}

To calculate $\sigma _{xy}^{I}$ we substitute $\tau $ from Eq. (\ref{tau-III}%
) 
into the Eq. (\ref{sIN}). First, from the formula (\ref{sxxDrude}) we find
the nonoscillating Hall conductivity, corresponding to the Drude model:%
\textbf{\ } 
\begin{equation}
	\overline{\sigma }_{xy}^{I}=-\omega _{c}\tau _{0}\overline{\sigma }_{xx}=-%
	\frac{e^{2}}{4\hbar d}\frac{\overline{\alpha }}{\gamma _{0}^{2}+\pi ^{2}}=%
	\frac{-\omega _{c}\tau _{0}\sigma _{0}}{1+\left( \omega _{c}\tau _{0}\right)
		^{2}}.  \label{sIxy0}
\end{equation}%
As we show below, $\overline{\sigma }_{xy}^{II}=0$, and the total
nonoscillating part $\overline{\sigma }_{xy}=\overline{\sigma }_{xy}^{I}$.
Substituting $\tau $ from Eq. (\ref{tau}) and Eq. (\ref{Sq}) into the Eq. (%
\ref{sIN}) we obtain 
\begin{gather}
	\sigma _{xy}^{I\,QO}\approx -2\omega _{c}\tau _{0}R_{D0}\cos \left( 
	\overline{\alpha }\right) J_{0}(\lambda )\overline{\sigma }_{xx}-\omega
	_{c}\tau _{0}\sigma _{xx}^{QO}  \notag \\
	\approx 2\overline{\sigma }_{xy}R_{D0}\left[ \frac{\gamma _{0}^{2}-\pi ^{2}}{%
		\gamma _{0}^{2}+\pi ^{2}}J_{0}\left( \lambda \right) \cos \left( \overline{%
		\alpha }\right) +\frac{\lambda }{\overline{\alpha }}J_{1}(\lambda )\sin (%
	\overline{\alpha })\right] .  \label{sIxy1}
\end{gather}%
The slow oscillating terms come from Eqs. (\ref{sIN}), (\ref{tau-III}), and (%
\ref{Se3}): 
\begin{equation}
	\sigma _{xy}^{SO}(\varepsilon )\approx \omega _{c}\tau _{0}\overline{\sigma
		_{xx}^{QO}\widetilde{\rho }/\rho _{0}}+\overline{\sigma }_{xy}\tau
	_{SO}/\tau _{0}-\omega _{c}\tau _{0}\sigma _{xx}^{SO},
	\label{sIxySO}
\end{equation}%
where the upper bar in the first term means averaging over the period of
MQO. As we show below in Sec. \ref{SecSxy}, $\sigma _{xy}^{II}$ does not
contribute to SO, because the DoS does not have slowly oscillating
term. Hence, $\sigma _{xy}^{SO}=\sigma _{xy}^{I\,SO}$. Substituting Eqs. (%
\ref{TildeRho}), (\ref{Sq}), (\ref{tau-III}) and (\ref{Ssl}) into (\ref%
{sIxySO}), we obtain 
\begin{gather}
	\sigma _{xy}^{SO}(\varepsilon )\approx 2\overline{\sigma }%
	_{xy}R_{D0}^{2}J_{0}^{2}\left( \lambda \right) \frac{\gamma _{0}^{2}(\gamma
		_{0}^{2}-3\pi ^{2})}{(\gamma _{0}^{2}+\pi ^{2})^{2}}  \label{sxySOg0} \\
	=2\overline{\sigma }_{xy}R_{D0}^{2}J_{0}^{2}\left( \lambda \right) \frac{%
		1-3(\omega _{c}\tau _{0})^{2}}{\left[ 1+(\omega _{c}\tau _{0})^{2}\right]
		^{2}}.  \label{sxySO1}
\end{gather}%
%
%
Eqs. (\ref{sxySOg0}), (\ref{sxySO1}) show that the amplitude of SO of the
Hall conductivity vanishes at $\gamma _{0}=\sqrt{3}\pi $ or $\omega _{c}\tau
_{0}=1/\sqrt{3}$.

\subsection{Calculation of $\protect\sigma _{xy}^{II}$}

\label{SecSxy}

The total electron density is (see Appendix \ref{App3}) 
\begin{gather}
	N(\mu )=\int_{0}^{\mu }\rho (B,\,\varepsilon )d\varepsilon \approx\rho _{0}
	\left[\mu +\frac{e\hbar B}{\pi m_*c} R_{D0}J_{0}(\lambda )\sin \left( 
	\overline{\alpha }\right) \right] ,  \label{Ntot0}
\end{gather}%
where we have substituted Eq. (\ref{rho}). It is not evident that the lowest
integration limit in Eq. (\ref{Ntot0}) is $0$, because the lowest Landau
level contributes the electron DoS $\rho (B,\,\varepsilon)$ even at $%
\varepsilon <0$ if $t_z>\hbar\omega_c/4$. The result (\ref{Ntot0}) is very
sensitive to this integration limit. The equivalence of Eq. (\ref{Ntot0}) 
to the initial formula for the DoS given by Eq. (\ref{rhon}) is shown in
Appendix \ref{App3}. Eqs. (\ref{sIIN}) and (\ref{Ntot0}) give 
\begin{equation}
	\sigma _{xy}^{II}=ec\frac{\partial N(\varepsilon )}{\partial B}\approx\rho _{0}%
	\frac{e^{2}\hbar }{\pi m_{\ast }}\frac{\partial }{dB}\left[
	BR_{D0}J_{0}(\lambda )\sin \left( \overline{\alpha }\right) \right] \text{.}
	\label{sIIder}
\end{equation}%
Note that $\sigma _{xy}^{II}$ does not contain nonoscillating terms. The are
also no SO in $\sigma _{xy}^{II}$, because there are no SO terms in the DoS,
and they do not appear in $N(\mu )$ in Eq. (\ref{Ntot0}). Using the relations%
\begin{gather}
	\frac{\partial R_{D0}}{\partial B}=R_{D0}\frac{\gamma_0}{B},  \label{A} \\
	\frac{\partial J_{0}(\lambda )}{\partial B}=-J_{1}(\lambda )\frac{\partial
		\lambda }{\partial B}=J_{1}(\lambda )\frac{\lambda }{B}, \\
	\frac{\partial \sin \left( \overline{\alpha }\right) }{\partial B}=
	-\cos (\overline{\alpha })\frac{\overline{\alpha }}{B},
\end{gather}%
we find from Eq. (\ref{sIIder}) 
\begin{gather}
	\sigma _{xy}^{II}\approx\rho _{0}\frac{e^{2}\hbar }{\pi m_{\ast }}R_{D0}\left[ -%
	\overline{\alpha }J_{0}(\lambda )\cos (\overline{\alpha })\right.
	\notag \\
	\left. +\lambda J_{1}(\lambda )\sin \left( \overline{\alpha }\right)
	+J_{0}(\lambda )(\gamma _{0}+1)\cos (\overline{\alpha })\right] .
	\label{sigma-xy-alpha}
\end{gather}%
Using Eq. (\ref{sIxy0}) and $\rho _{0}=m_{\ast }/(2\pi \hbar ^{2}d)$ we
rewrite Eq. (\ref{sigma-xy-alpha}) as 
\begin{gather}
	\sigma _{xy}^{II}\approx2\overline{\sigma }_{xy}R_{D0}\frac{\gamma _{0}^{2}+\pi
		^{2}}{\pi ^{2}}\left[ J_{0}(\lambda )\cos (\overline{\alpha })\right.
	\label{sIIxy1} \\
	\left. -\overline{\alpha }^{-1}\left[ \lambda J_{1}(\lambda )\sin \left( 
	\overline{\alpha }\right) +J_{0}(\lambda )(\gamma _{0}+1)\cos (\overline{%
		\alpha })\right] \right] .  \notag
\end{gather}

\subsection{Total MQO of Hall conductivity}

Combining Eqs. (\ref{sIxy1}) and (\ref{sIIxy1}) we obtain the MQO of Hall
conductivity%
\begin{equation}
	\sigma _{xy}^{QO}=\sigma _{xy}^{I\,QO}+\sigma _{xy}^{II}=\widetilde{\sigma }%
	_{xy}^{QO}+\Delta _{\lambda }\sigma _{xy}^{QO}+\Delta _{\gamma }%
\sigma _{xy}^{QO}.  \label{sxyQO}
\end{equation}%
At $\overline{\alpha }\gg \lambda ,\,\gamma _{0},\,1$, corresponding to $\mu
\gg t_{z},\,\hbar /\tau ,\,\hbar \omega _{c}$, one may keep only the first
main term in the square brackets of Eqs. (\ref{sIxy1}) and (\ref{sIIxy1}),
which gives the dominant 
$\widetilde{\sigma }_{xy}^{QO}$ 
MQO term in the leading order in $%
\overline{\alpha }$:%
\begin{gather}
	\widetilde{\sigma }_{xy}^{QO}\approx2\overline{\sigma }_{xy}R_{D0}J_{0}\left(
	\lambda \right) \cos \left( \overline{\alpha }\right) \left[ \frac{\gamma
		_{0}^{2}-\pi ^{2}}{\gamma _{0}^{2}+\pi ^{2}}+\frac{\gamma _{0}^{2}+\pi ^{2}}{%
		\pi ^{2}}\right]   \notag \\
	=2\overline{\sigma }_{xy}R_{D0}J_{0}\left( \lambda \right) \cos \left( 
	\overline{\alpha }\right) \frac{\gamma _{0}^{2}}{\pi ^{2}}\frac{\gamma
		_{0}^{2}+3\pi ^{2}}{\gamma _{0}^{2}+\pi ^{2}}  \label{sxyQOmain} \\
	=-\frac{\overline{\sigma }_{xy}}{\left( \omega _{c}\tau _{0}\right) ^{2}}%
	\frac{1+3\left( \omega _{c}\tau _{0}\right) ^{2}}{1+\left( \omega _{c}\tau
		_{0}\right) ^{2}}\frac{\widetilde{\rho }}{\rho _{0}}.  \label{sxyQOmain1}
\end{gather}%
This formula coincides with Eq. (3.25) of Ref. \cite{Isihara1986Dec}, but
the oscillating DoS now contains an extra factor $J_{0}(\lambda )$. The
remaining two terms in Eq. (\ref{sxyQO}) are smaller by the factors $\sim
t_{z}/\mu $ and $\sim \hbar \omega _{c}/\mu $: 
\begin{gather}
	\Delta _{\lambda }\sigma _{xy}^{QO} \approx2\overline{\sigma }%
	_{xy}R_{D0}\frac{\lambda }{\overline{\alpha }}J_{1}(\lambda )\sin (\overline{%
		\alpha })\left[ 1-\frac{\gamma _{0}^{2}+\pi ^{2}}{\pi ^{2}}\right]   \notag
	\\
	=-2\overline{\sigma }_{xy}R_{D0}\frac{\lambda }{\overline{\alpha }}\frac{%
		\gamma _{0}^{2}}{\pi ^{2}}J_{1}(\lambda )\sin (\overline{\alpha })  \notag \\
	=-4\overline{\sigma }_{xy}\frac{R_{D0}}{\left( \omega _{c}\tau \right) ^{2}%
	}\frac{t_{z}}{\mu}J_{1}(\lambda )\sin (\overline{\alpha })
	\label{sxyL}
\end{gather}%
and 
\begin{gather}
	\Delta _{\gamma }\sigma _{xy}^{QO}\approx -2\overline{\sigma }%
	_{xy}R_{D0}J_{0}(\lambda )\frac{\gamma^2_0+\pi^2}{\pi^2}\frac{\gamma _{0}+1}{\overline{\alpha }}\cos (%
	\overline{\alpha })\notag\\\approx\overline{\sigma }_{xy}\frac{\widetilde{\rho }}{\rho
		_{0}}\frac{\gamma^2_0+\pi^2}{\pi^2}\frac{\gamma _{0}+1}{\overline{\alpha }}.  \label{sxyQOAdd}
\end{gather}%
The term $\Delta _{\lambda }\sigma _{xy}^{QO}$ leads to the
phase shift of MQO and to a non-zero MQO amplitude in the beat nodes,
similar to its effect in $\widetilde{\sigma }_{xx}^{QO}$ given by Eqs.
(39)-(42) of Ref. \cite{Mogilyuk2018Jul}. It may be notable in various
compounds where the $k_{z}$ dispersion amplitude $4t_{z}$ is comparable to
the Fermi energy. The term $\Delta _{\gamma }\sigma _{xy}^{QO}$
slightly changes the amplitude of MQO and is only important in semimetals.

At finite $T$ after the averaging over a thermodynamic ensemble Eqs. (\ref%
{sxyQOmain1})-(\ref{sxyQOAdd}) are multiplied by the temperature damping
factor $R_T$ given by Eq. (\ref{RT}). Eqs. (\ref{sIxy0}), (\ref{sxySO1}) and (%
\ref{sxyQO})-(\ref{sxyQOAdd}) for $\sigma_{xy}$, together with Eqs. (\ref%
{Se3})-(\ref{sxxSlO}) for $\sigma_{xx}=\sigma_{yy}$, complete our
calculations of the magnetic oscillations of the conductivity tensor and
allow to find the resistivity tensor. 

\begin{figure}[ht]
	\centering
	\includegraphics[width=0.5\textwidth]{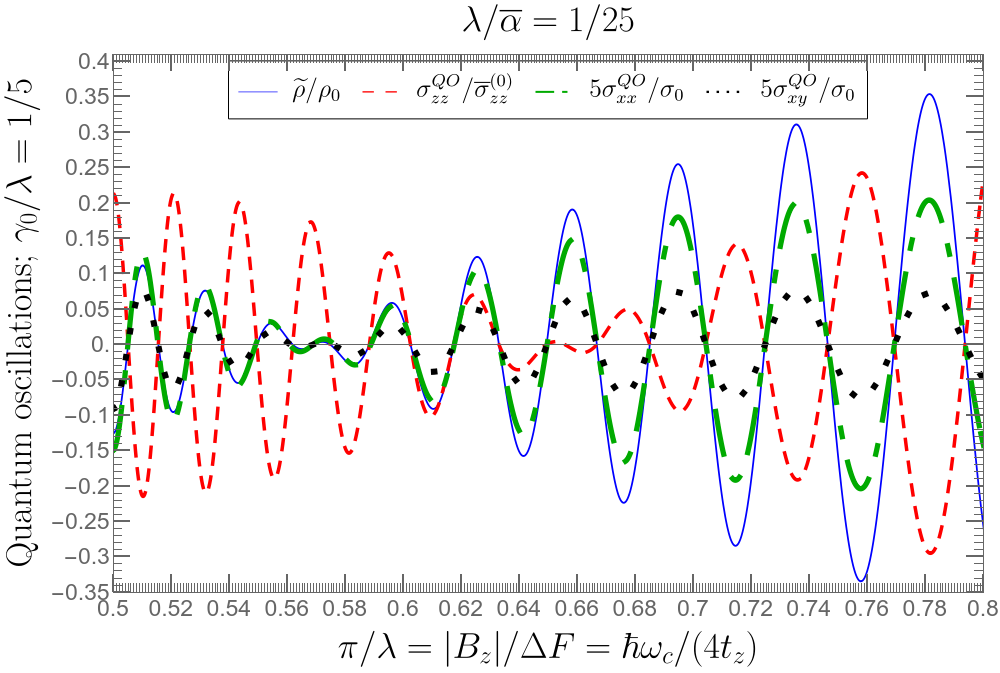}
\caption{Magnetic quantum oscillations at zero temperature of the density of states $\widetilde{\rho}$, interlayer conductivity $\sigma^{QO}_{zz}$, diagonal conductivity $\sigma^{QO}_{xx}$ and Hall conductivity $\sigma^{QO}_{xy}$ at $\gamma_{0}/\lambda=1/5$ and $\lambda/\overline{\alpha}=1/25$. It can be seen that the oscillations of $\widetilde{\rho}$, $\sigma^{QO}_{xx}$, $\sigma^{QO}_{xy}$ in the selected range of magnetic field are in phase and have the same beat-node positions. Meanwhile, the oscillations of $\sigma^{QO}_{zz}$ and $\widetilde{\rho}$ are mostly in antiphase except for the short interval $|B_z|/\Delta F\in(0.57,\, 0.66)$ between the beat nodes of $\sigma^{QO}_{xx}$ and $\sigma^{QO}_{zz}$, because the latter is shifted.}
\label{FigMQOsigma}
\end{figure}

\begin{figure}[ht]
\centering
\includegraphics[width=0.5\textwidth]{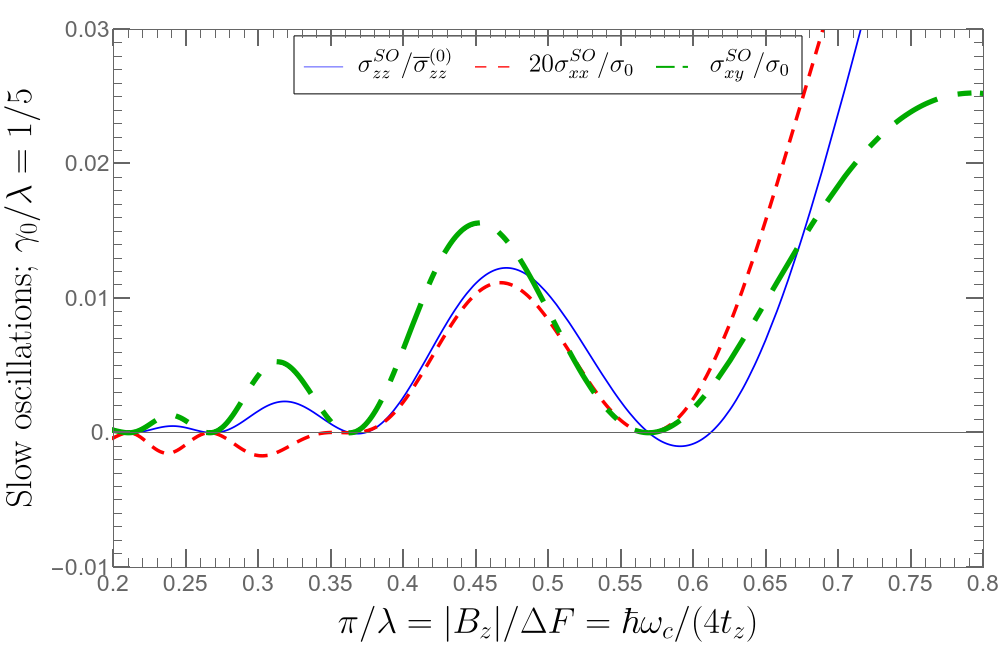}
\caption{Slow oscillations of interlayer conductivity $\sigma^{SO}_{zz}$, diagonal conductivity $\sigma^{SO}_{xx}$ and Hall conductivity $\sigma^{SO}_{xy}$ at $\gamma_{0}/\lambda=1/5$. The SO amplitude of diagonal in-plane conductivity $\sigma^{SO}_{xx}$ changes its sign with phase inversion at $|B_z|/\Delta F\approx0.35$. The SO of $\sigma_{zz}$ are slightly shifted in phase according to Eq. (\ref{eq:szzSO}).}
\label{FigSOsigma}
\end{figure}

In Figs. \ref{FigMQOsigma} and \ref{FigSOsigma} we compare the calculated MQO and SO of the in-plane Hall and diagonal components of conductivity tensor, given by Eqs. (\ref{sxyQO})-(\ref{sxyQOAdd}), (\ref{sxySO1}) for $\sigma_{xy}$ and Eqs. (\ref%
{Se3})-(\ref{sxxSlO}) for $\sigma_{xx}$, with its interlayer component given by Eqs. (\ref{szz})-(\ref{eq:szzSO}) and with the MQO of DoS given by Eq. (\ref{TildeRho}). The parameters for the plot in Fig. \ref{FigMQOsigma} are indicated in the figure and chosen to illustrate the behavior near the beat node. The MQO beat-node positions of $\sigma_{xx},\,\sigma_{xy}$ and $\widetilde{\rho }/\rho _{0}$ coincide, while the beat-node position in $\sigma_{zz}$ is shifted. The parameters for the plot in Fig. \ref{FigSOsigma} are also indicated in the plot and chosen to illustrate the zero and sign change of the SO amplitude of $\sigma_{xx}$ at $\omega _{c}\tau _{0} \approx  \sqrt{3}$, corresponding to $|B_z|/\Delta F = \omega _{c}\tau _{0}\,\gamma_{0}/\lambda \approx 0.35$ in Fig. \ref{FigSOsigma}.

\section{Calculation of resistivity tensor}\label{SecR}

The resistivity tensor $\hat{R}\equiv \hat{\sigma}^{-1}$ can be computed
directly by the inversion of conductivity tensor $\hat{\sigma}$ calculated
in the previous section. However, a question appears about the correct order
of two operations: the inversion of conductivity tensor and the temperature
smearing. While this order does not affect the amplitude of the first MQO
harmonic, it changes the higher harmonics and the differential slow
oscillations. The correct order is not obvious and may depend on the
physical system and even on the experimental setup. The same question about
the temperature smearing of MQO appears for the longitudinal interlayer
magnetoresistance $R_{zz}\left( B_{z}\right) =\sigma _{zz}^{-1}\left(
B_{z}\right) $, as shown below. 

\subsection{Interlayer resistivity and its temperature averaging} \label{SubSecRzz}

The calculation of interlayer magnetoresistance $R_{zz}\left( B_{z}\right) $ does not require inverting the
in-plane conductivity tensor and is simpler than that of the in-plane magnetoresistance $R_{xx}\left( B_{z}\right) $
or $R_{xy}\left( B_{z}\right) $. The interlayer magnetoconductivity in quasi-2D
metals with SO is given by \cite{Kartsovnik2002Aug,Grigoriev2003Apr,Mogilyuk2018Jul} 
\begin{equation}
	\sigma _{zz}(\mu,\,B_{z})=\overline{\sigma }_{zz}^{(0)}+\sigma _{zz}^{QO}(\mu
	)+\sigma _{zz}^{SO}(\mu ).  \label{szz}
\end{equation}%
Here the monotonic part $\overline{\sigma }_{zz}^{(0)}\left(B_{z}\right) $
depends on $B_{z}$ only in a high magnetic field at $\hbar\omega_{c}>\max
\{4t_{z},\,\pi\hbar/\left( 2\tau _{0}\right)\}$ when the SO are absent, and at 
$\hbar\omega_{c}\gg 4t_{z},\,\hbar/\tau _{0}$ it has the asymptotic behavior $\overline{\sigma}%
_{zz}^{(0)}\left( B_{z}\right) \propto B_{z}^{-1/2}$ according to Refs. \cite%
{Grigoriev2011Jun,Grigoriev2011Sep,Grigoriev2012Oct,Grigoriev2013Aug,Grigoriev2014Sep}. 
The MQO of $\sigma _{zz}=\sigma _{zz}(\mu,\,B_{z})$ in the first order of the Dingle factor $R_{D}$
are given by \cite{Grigoriev2003Apr,Mogilyuk2018Jul}%
\begin{gather}
	\sigma _{zz}^{QO}(\mu,\,B_{z})\approx 2\overline{\sigma }_{zz}^{(0)}\cos \left( 
	\overline{\alpha }\right) R_{D0}R_{T}R_{W}  \label{eq:szzqo} \\
	\times \left[ J_{0}\left( \lambda \right) -\frac{2}{\lambda }\left(
	1+\gamma _{0}\right) J_{1}\left( \lambda \right) \right] ,  \notag
\end{gather}%
where, by the analogy with Ref. \cite{Grigoriev2017Oct}, we add the factor $%
R_{W}$ to describe the MQO damping by the phase smearing due to long-range
spatial inhomogeneities \cite%
{Kartsovnik2002Aug,Grigoriev2012Oct,Grigoriev2017Oct}. The SO
of $\sigma _{zz}$ are given by Eq. (18) of Ref. \cite%
{Grigoriev2003Apr} or Eq. (64) of Ref. \cite{Mogilyuk2018Jul}: 
\begin{equation}
	\sigma _{zz}^{SO}(\mu,\,B_{z})\approx 2\overline{\sigma }_{zz}^{(0)}R_{D0}^{2}J_{0}
	\left( \lambda \right) \left[ J_{0}\left( \lambda \right) -\frac{2}{\lambda }
	J_{1}\left( \lambda \right) \right] .  \label{eq:szzSO}
\end{equation}
The interlayer resistivity in the lowest order in the Dingle factor is also given
by the sum of three terms, 
\begin{equation}
	R_{zz}=\sigma _{zz}^{-1}=\overline{R}_{zz}+R_{zz}^{QO}+R_{zz}^{SO},  \label{Rzz}
\end{equation}%
where the monotonic term $\overline{R}_{zz}=1/\overline{\sigma }_{zz}^{(0)}$ and
the MQO of resistivity 
\begin{gather}
	R_{zz}^{QO}(\mu )\approx -2\overline{R}_{zz}\cos \left( \overline{\alpha }%
	\right) R_{D0}R_{T}R_{W}  \label{RQO} \\
	\times \left[ J_{0}\left( \lambda \right) -\frac{2}{\lambda }\left(
	1+\gamma _{0}\right) J_{1}\left( \lambda \right) \right]   \notag
\end{gather}%
do not depend on the order of temperature smearing and of conductivity
inversion. Only the second- and higher-order terms in $R_{D}$, including the
term $R_{zz}^{SO}$ describing the differential oscillations, depend on the
order of these two operations.

If during the calculation of interlayer resistivity the temperature smearing
is applied to the conductivity before its inversion (\emph{first method or} $%
\sigma $\emph{-averaging}), we obtain in the lowest second order in $R_{D}$
the following SO term:%
\begin{gather}
	R_{zz}^{SO}/\overline{R}_{zz}\approx \left( \sigma _{zz}^{QO}/\overline{\sigma }
	_{zz}^{(0)}\right) ^{2}-\sigma _{zz}^{SO}/\overline{\sigma }_{zz}^{(0)} \\
	=-2R_{D0}^{2}J_{0}\left( \lambda \right) \left[ J_{0}\left( \lambda \right) - 
	\frac{2}{\lambda }J_{1}\left( \lambda \right) \right] +  \label{RzzSav} \\
	+2R_{D0}^{2}\left( R_{T}R_{W}\right) ^{2}\left[ J_{0}\left( \lambda \right) - 
	\frac{2}{\lambda }\left( 1+\gamma _{0}\right) J_{1}\left( \lambda \right) %
	\right] ^{2}.  \notag
\end{gather}%
If in the last line we assume $\left( R_{T}R_{W}\right) ^{2}\ll 1$, as it
usually happens \cite{Kartsovnik2002Aug}, we obtain 
\begin{equation}
	R_{zz}^{SO}/\overline{R}_{zz}\approx -2R_{D0}^{2}J_{0}\left( \lambda \right) \left[
	J_{0}\left( \lambda \right) -\frac{2}{\lambda }J_{1}\left( \lambda \right) %
	\right] .  \label{RzzSO1}
\end{equation}%
It is this expression after the large-argument asymptotic expansion of the Bessel functions was used to analyze the experimental data in Ref. \cite{Kartsovnik2002Aug} and shown to be reasonably consistent.

If during the calculation of interlayer resistivity in Eq. (\ref{Rzz}) the
smearing over temperature and sample inhomogeneities is applied to
resistivity, i.e. only on the final step after the conductivity inversion (\emph{second method or
	R-averaging}), we obtain%
\begin{gather}
	R_{zz}^{SO}/\overline{R}_{zz}\approx-2R_{D0}^{2}J_{0}\left( \lambda \right) \left[
	J_{0}\left( \lambda \right) -\frac{2}{\lambda }J_{1}\left( \lambda \right) %
	\right] +  \notag \\
	+2R_{D0}^{2}\left[ J_{0}\left( \lambda \right) -\frac{2}{\lambda }\left(
	1+\gamma _{0}\right) J_{1}\left( \lambda \right) \right] ^{2}  \notag \\
	\approx -\frac{8}{\lambda }R_{D0}^{2}J_{1}\left( \lambda \right) \left[
	\left( 1/2+\gamma _{0}\right) J_{0}\left( \lambda \right) -J_{1}\left(
	\lambda \right) \left( 1+\gamma _{0}\right) ^{2}/\lambda \right] .
	\label{RzzSO2}
\end{gather}%
The SO described by Eq. (\ref{RzzSO2}) are smaller than those in Eq. (\ref%
{RzzSO1}) by a factor $\sim 2/\lambda $ and have a different phase. The
usual Shubnikov oscillations $R_{zz}^{QO}$, given by Eq. (\ref{RQO}), in the
lowest main order in $R_{D0}$ do not depend on the sequence of temperature
averaging and conductivity inversion. Hence, by the comparison of the phase
of SO and of the beats of Shubnikov oscillations one can experimentally
determine what type of temperature averaging one should apply for the
particular physical case. Unfortunately, in Ref. \cite{Kartsovnik2002Aug} 
no comparison of SO phase with the phase of MQO beats was performed.

Besides the temperature, the long-range (macroscopic) sample inhomogeneities
result to the local variations of the Fermi energy $\mu $ and of the MQO
phase \cite{Kartsovnik2002Aug,Grigoriev2012Oct,Grigoriev2017Oct}. They give
a major contribution to the MQO damping, described by the factor $R_{W}$\ in
Eq. (\ref{eq:szzqo}), and to the Dingle temperature in organic metals, as
observed by the comparison of the Dingle factors of usual MQO (fast oscillations) and
differential SO (slow oscillations) \cite{Kartsovnik2002Aug}.
Since the SO are not damped by such macroscopic inhomogeneities, they are
often much stronger than the usual MQO \cite{Kartsovnik2002Aug,Grigoriev2017Oct}, 
because the Dingle temperature of SO is considerably smaller than that of MQO. 
The macroscopic Fermi-energy inhomogeneities also result to the Gaussian
shape of Landau levels and to the strong modification of the field and
harmonic dependence of the observed Dingle factor of MQO \cite%
{Grigoriev2012Oct}. Averaging of resistivity or conductivity MQO over these
inhomogeneities corresponds to their serial and parallel connection
correspondingly. In real crystals these inhomogeneities appear in all
directions, both parallel and perpendicular to the electric current. Hence,
to describe these inhomogeneities it may be correct to average neither
resistivity nor conductivity but an intermediate quantity, corresponding 
to an effective conductivity of a heterogeneous media \cite{Torquato2002}.

A related question emerged in Ref. \cite{Grigoriev2012Oct}, where the
measured field dependence of interlayer magnetoresistance $R_{zz}\left(
B_{z}\right) $ in the quasi-2D organic metal $\alpha$-(BEDT-TTF)$_{2}$%
KHg(SCN)$_{4}$ was compared to the theoretical prediction \cite%
{Grigoriev2011Jun,Grigoriev2011Sep} for its monotonic part $\overline{R}%
_{zz}\left( B_{z}\right) 
\propto \sqrt{B_{z}}$. 
In a high magnetic field the experimental curve $R_{zz}\left( B_{z}\right) $
contains strong MQO. Its averaging over the MQO period to extract the
monotonic part $\overline{R}_{zz}\left( B_{z}\right) $ can be performed in two
ways, depending on whether the resistance $R_{zz}$ or conductivity $\sigma
_{zz}=R_{zz}^{-1}$ must be averaged. 
This
example shows that in addition to the method of averaging of theoretical
results, described by Eqs. (\ref{RzzSav})-(\ref{RzzSO2}), the averaging of
experimental data over the MQO period is also important for their correct
comparison with theory.

\subsection{In-plane resistivity tensor}

As we have shown above, the question about the correct way of temperature
averaging can be answered experimentally for any particular system. Below we
give the result for both methods of temperature averaging. The resistivity
tensor $\hat{R}\equiv\hat{\sigma}^{-1}$ can be calculated directly using
the results of previous section. Similar to Eq. (\ref{sxyQO}), we separate
the main oscillating term $\hat{\sigma}_{m}$ and the small correction $%
\Delta _{\lambda }\hat{\sigma}^{QO}\sim \left( 2t_{z}/\mu\right) \hat{\sigma}$.

The main terms in Eqs. (\ref{Se3})-(\ref{sxxSlO}) for $\sigma _{xx}=\sigma
_{yy}$ together with Eqs. (\ref{sIxy0}), (\ref{sxySO1}) and (\ref{sxyQO})-(%
\ref{sxyQOmain1}) for $\sigma _{xy}=-\sigma _{xy}$ give the following
in-plane conductivity tensor 
\begin{equation}
	\frac{\hat{\sigma}_{m}}{\overline{\sigma }_{xx}}=\left( 
	\begin{array}{cc}
		A_{xx} & A_{xy} \\ 
		A_{yx} & A_{yy}%
	\end{array}%
	\right) ,  \label{sm}
\end{equation}%
where the diagonal term is%
\begin{gather}
	A_{xx}\approx 1+\frac{2\left( \omega _{c}\tau _{0}\right) ^{2}}{1+\left(
		\omega _{c}\tau _{0}\right) ^{2}}\frac{\widetilde{\rho }}{\rho _{0}}%
	R_{T}R_{W}  \label{AxxT} \\
	+2R_{D0}^{2}J_{0}^{2}(\lambda )\frac{\left( \omega _{c}\tau _{0}\right) ^{2}%
		\left[ \left( \omega _{c}\tau _{0}\right) ^{2}-3\right] }{\left[ 1+\left(
		\omega _{c}\tau _{0}\right) ^{2}\right] ^{2}},  \notag
\end{gather}
the off-diagonal term is%
\begin{gather}
	A_{xy}\approx -\omega _{c}\tau _{0}\left( 1-\frac{R_{T}R_{W}}{\left( \omega
		_{c}\tau _{0}\right) ^{2}}\frac{1+3\left( \omega _{c}\tau _{0}\right) ^{2}}{%
		1+\left( \omega _{c}\tau _{0}\right) ^{2}}\frac{\widetilde{\rho }}{\rho _{0}}%
	\right.   \notag \\
	\left. +\frac{2R_{D0}^{2}J_{0}^{2}(\lambda )(1-3(\omega _{c}\tau _{0})^{2})}{%
		[1+(\omega _{c}\tau _{0})^{2}]^{2}}\right) ,  \label{AxyT}
\end{gather}%
and the conductivity tensor has the following symmetry properties: 
\begin{equation}
	A_{xy}=-A_{yx},\,A_{yy}=A_{xx}.  \label{SymA}
\end{equation}

\begin{figure}[th]
	\centering
	\includegraphics[width=0.5\textwidth]{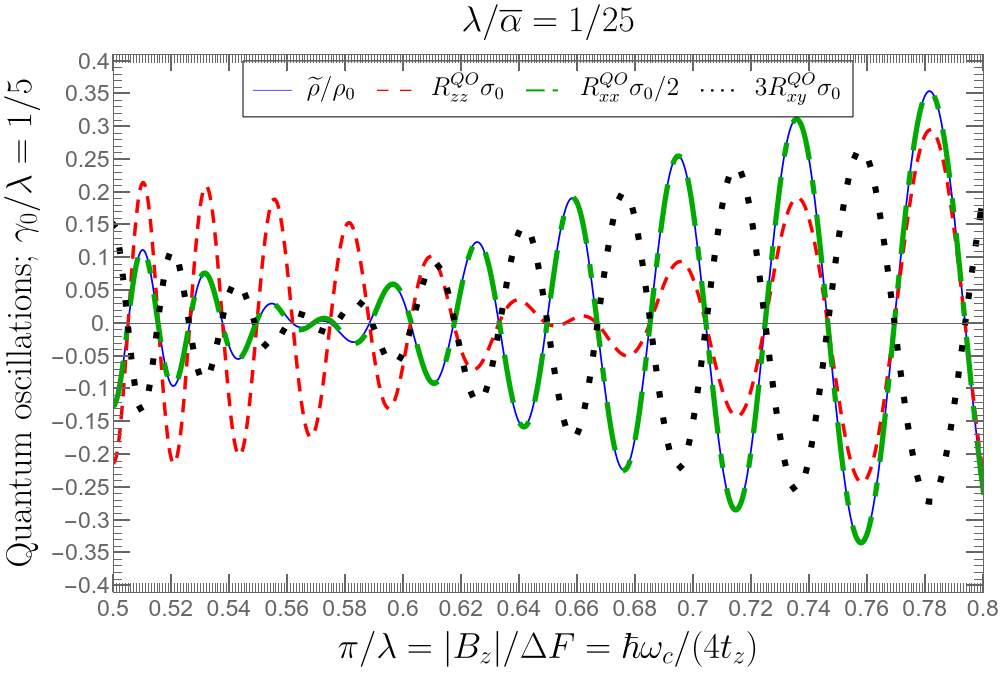}
\caption{Magnetic quantum oscillations at zero temperature of the density of states $\widetilde{\rho}$, interlayer resistivity $R^{QO}_{zz}$, diagonal resistivity  $R^{QO}_{xx}$ and Hall resistivity  $R^{QO}_{xy}$ at $\gamma_{0}/\lambda=1/5$ and $\lambda/\overline{\alpha}=1/25$. The oscillations of $\widetilde{\rho}$ and $R^{QO}_{xx}$ are always in phase, while the $R^{QO}_{xy}$ oscillations are antiphase with them. The oscillations of $R^{QO}_{zz}$ are mostly in phase with $\widetilde{\rho}$ and $R^{QO}_{xx}$ except for the short interval $|B_z|/\Delta F\in(0.56,\, 0.67)$ near the beat node which is shifted for $R^{QO}_{zz}$.} 
\label{FigRMQO}
\end{figure}

The formal inversion of the conductivity tensor in Eq. (\ref{sm}) gives the
following resistivity tensor 
\begin{equation}
	\hat{R}\equiv \hat{\sigma}^{-1}=\frac{1}{\sigma _{0}}\left( 
	\begin{array}{cc}
		B_{xx} & B_{xy} \\ 
		B_{yx} & B_{yy}%
	\end{array}%
	\right) ,\,\text{where}  \label{R}
\end{equation}%
\begin{gather}
	B_{xx}\approx 1+2\frac{\widetilde{\rho }}{\rho _{0}}R_{T}R_{W}+\frac{%
		2R_{D0}^{2}J_{0}^{2}(\lambda )\left[ (\omega _{c}\tau
		_{0})^{4}-R_{T}^{2}R_{W}^{2}\right] }{(\omega _{c}\tau _{0})^{2}\left[
		1+(\omega _{c}\tau _{0})^{2}\right] },  \label{RtensorXX} \\
	B_{xy}\approx \omega _{c}\tau _{0}-\frac{\widetilde{\rho }R_{T}R_{W}}{\omega
		_{c}\tau _{0}\rho _{0}}+  \notag \\
	\frac{2R_{D0}^{2}J_{0}^{2}(\lambda )\left[ (\omega _{c}\tau
		_{0})^{2}-R_{T}^{2}R_{W}^{2}\left[ 3+4(\omega _{c}\tau _{0})^{2}\right] %
		\right] }{\omega _{c}\tau _{0}\left[ 1+(\omega _{c}\tau _{0})^{2}\right] },
	\label{RtensorXY} \\
	B_{xy}=-B_{yx},\,B_{yy}=B_{xx}.  \label{Rtensor0}
\end{gather}%
Here we expanded the result in the powers of $R_{D0}$, and in the second order
in $R_{D}$ we only kept the SO terms.
 According to these formulas, the MQO of diagonal and Hall magnetoresistance 
components, given by the second terms in Eqs. (\ref{RtensorXX}) and (\ref{RtensorXY}), 
have opposite phase. This does not depend on the averaging procedure, 
because the MQO appear in the first order of the Dingle factor. The resistivity MQO given by Eqs. (\ref{RQO}) and (\ref{RxxMQO})-(\ref{PhiXY}) are shown in Figs. \ref{FigRMQO} and \ref{RMQOF6}. The SO of 
diagonal and Hall magnetoresistance components may have the same or opposite phase, 
depending on the averaging procedure, on magnetic field strength, and on damping factors $R_T R_W$.

\subsubsection{Resistivity tensor for $\protect\sigma $-averaging}

For $\protect\sigma $-averaging the resistivity tensor is given by Eqs. (\ref{R})-(\ref{Rtensor0}), where the damping factors $R_T$ and $R_W$ in the 
SO terms of Eqs. (\ref{RtensorXX}) and (\ref{RtensorXY}) are the same as in the second MQO terms.
Hence, for this type of averaging the mutual phase and the amplitudes of $R_{xx}$ and $R_{xy}$ 
SO can be controlled by temperature and magnetic field strength.
Unlike $\sigma _{xx}$ and $\sigma _{xy}$ SO oscillations, given by Eqs. (\ref%
{sxxSlO}) and (\ref{sxySO1}), the zero point where the amplitude of resistivity
SO in Eqs. (\ref{RtensorXX}) and (\ref{RtensorXY}) vanishes, for $\sigma $\emph{-averaging} depends on temperature and on 
spatial heterogeneity via the suppression factor $R_{T}R_{W}$. 

According to Eq. (\ref{RtensorXX}), the amplitude of the SO of diagonal 
components $R_{xx}=R_{yy}$ vanishes at 
\begin{equation}
	\omega _{c}\tau _{0}=\sqrt{R_{T}R_{W}}.  \label{RxxSOVanish}
\end{equation}%
This condition is rarely satisfied at any magnetic field $B_{z}$, because
the product $R_{T}R_{W}$ increases from an exponentially small value at $%
\omega _{c}\tau _{0}\ll 1$ to unity at $\omega _{c}\tau _{0}\gg 1$, and
usually $R_{T}R_{W}\lesssim R_{D0}=\exp \left( -\pi /(\omega _{c}\tau
_{0})\right) \ll \left( \omega _{c}\tau _{0}\right) ^{2}$ at all $B_{z}$.
Only if $R_{T}R_{W}>20R_{D0}$ at $\omega _{c}\tau _{0}\sim 1$ Eq. (\ref%
{RxxSOVanish}) can be satisfied, but, usually, $R_{T}R_{W}<R_{D0}$. 

The amplitude
of slow oscillations of non-diagonal components $\rho _{xy}^{R}=-\rho
_{yx}^{R}$ vanishes at some field $B_{z}$ corresponding to
\begin{equation}
	\omega _{c}\tau _{0}=R_{T}R_{W}\sqrt{3/(1-4R_{T}^{2}R_{W}^{2})}.
	\label{RxySOVanish}
\end{equation}
Formally, Eq. (\ref{RxySOVanish}) always has a real solution, because $%
R_{T}R_{W}$ depends on magnetic field $B_{z}$, and with the increase of $B_{z}$
the right hand side of this equation increase from an exponentially small
value at $\omega _{c}\tau _{0}\ll 1$ to infinity at $\omega _{c}\tau _{0}>1$
when $R_{T}R_{W}=1/2$. Nevertheless, if $R_{W}\ll 1$ at the available
magnetic field, the zero point of the SO amplitude of $R_{xy}$ may happen at
too high magnetic field, above the experimental window.

\begin{figure}[t]
	\centering
	\includegraphics[width=0.5\textwidth]{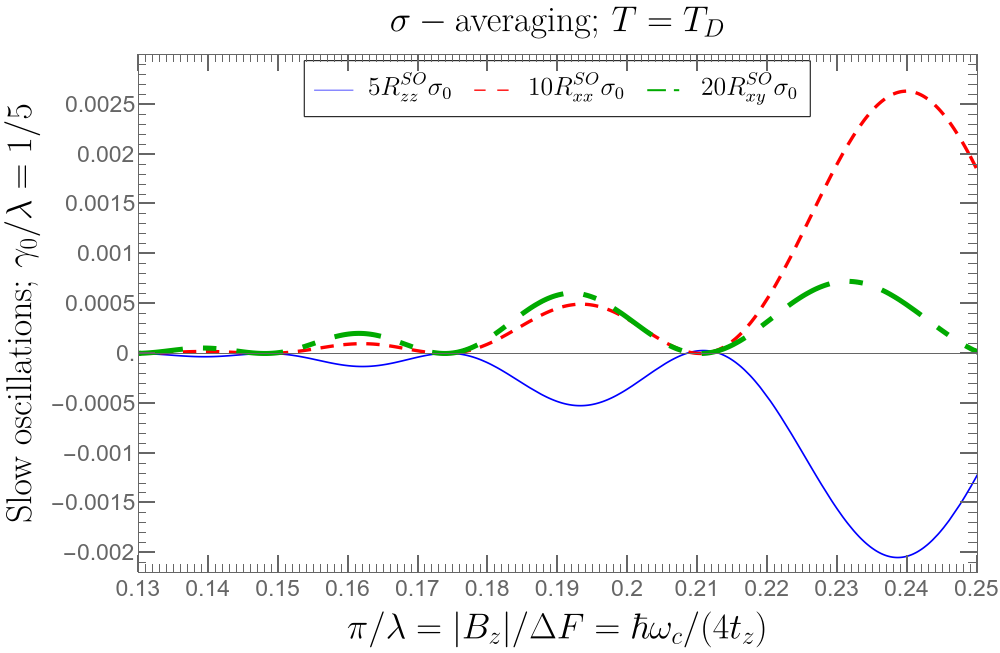}
\caption{Slow oscillations of interlayer resistivity $R^{SO}_{zz}$, diagonal resistivity $R^{SO}_{xx}$ and Hall resistivity $R^{SO}_{xy}$ , given by Eqs. (\ref{RzzSO1}) and (\ref{R})-(\ref{RtensorXY}) as a result of  $\sigma$-averaging at temperature $T=T_D$ and $\gamma_{0}/\lambda=1/5$. The slow oscillations $R^{SO}_{zz}$ and $R^{SO}_{xx}$ are antiphase, and the slow oscillations $R^{SO}_{xx}$ and $R^{SO}_{xy}$ are in phase.}
\label{FigSav}
\end{figure}

At $R_{T}R_{W}\ll 1$ Eqs. (\ref{RtensorXX}) and (\ref{RtensorXY}) simplify
to 
\begin{gather}
	B_{xx}\approx 1+2\frac{\widetilde{\rho }}{\rho _{0}}R_{T}R_{W}+\frac{%
		2R_{D0}^{2}J_{0}^{2}(\lambda )(\omega _{c}\tau _{0})^{2}}{1+(\omega _{c}\tau
		_{0})^{2}},  \label{BxxSav} \\
	B_{xy}\approx \omega _{c}\tau _{0}-\frac{\widetilde{\rho }R_{T}R_{W}}{\omega
		_{c}\tau _{0}\rho _{0}}+\frac{2\omega _{c}\tau
		_{0}R_{D0}^{2}J_{0}^{2}(\lambda )}{1+(\omega _{c}\tau _{0})^{2}}.
	\label{BxySav}
\end{gather}
In this limiting case the MQO of diagonal and Hall magnetoresistance 
components have opposite phase, and their SO have the same phase. Comparing Eqs. (\ref{BxxSav}) and (\ref{BxySav}) with Eq. (\ref{RzzSO1}) we note that at $R_{T}R_{W}\ll 1$ and $2/\lambda = \hbar\omega_c /(2\pi t_z) \ll 1$ the phase of SO of $R_{xx}$ and $R_{xy}$ is nearly opposite to the SO phase of $R_{zz}$. This case is illustrated in Fig. \ref{FigSav}.

\subsubsection{Resistivity tensor at $R$-averaging}

\begin{figure}[t]
\centering
\includegraphics[width=0.5\textwidth]{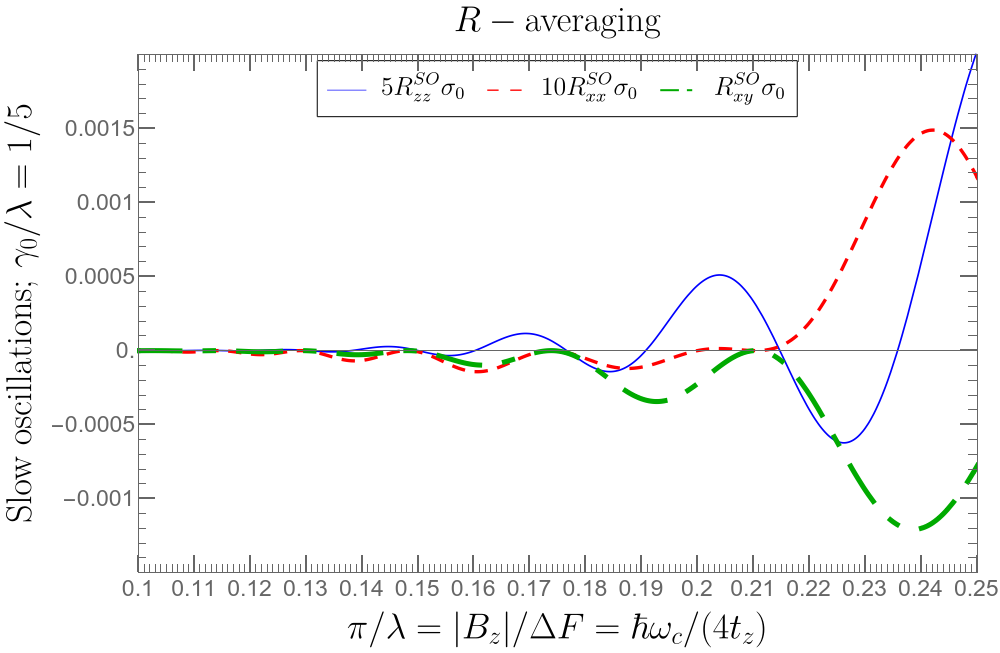}
\caption{Slow oscillations of interlayer resistivity $R^{SO}_{zz}$, diagonal resistivity $R^{SO}_{xx}$ and Hall resistivity $R^{SO}_{xy}$, given by Eqs. (\ref{RzzSO2}), (\ref{BxxRav}) and (\ref{BxyRav}) as a result of  $R$-averaging at $\gamma_{0}/\lambda=1/5$. Slow oscillations $R^{SO}_{zz}$ and $R^{SO}_{xx}$ are approximately shifted by a quarter of a period relative to each other. The SO oscillations $R^{SO}_{xx}$ and $R^{SO}_{xy}$ have the same phase in a weak magnetic field and opposite phases in a strong field, according to Eqs. (\ref{BxxRav}) and (\ref{BxyRav}).}
\label{FigRav}
\end{figure}

If the averaging over temperature and long-range spatial inhomogeneities is
applied only to resistivity at the final stage, the $R_{T}$ and $R_{W}$
factors in the conductivity tensor given by Eqs. (\ref{sm})-(\ref{AxyT})
should be replaced by unity before the tensor inversion. Hence, for $R$-averaging 
the resistivity tensor is again given by Eqs. (\ref{R})-(\ref{Rtensor0}), where the 
damping factors $R_T$ and $R_W$ appear only in the second MQO terms of Eqs. (\ref{RtensorXX}) 
and (\ref{RtensorXY}), while in the third SO terms of Eqs. (\ref{RtensorXX}) 
and (\ref{RtensorXY}) the factors $R_T$ and $R_W$ must be replaced by unity.
The leading term of the in-plane resistivity tensor is then given by the matrix in
Eq. (\ref{R}) with the elements   
\begin{gather}
	B_{xx}\approx 1+2\frac{\widetilde{\rho }}{\rho _{0}}%
	R_{T}R_{W}+2R_{D0}^{2}J_{0}^{2}(\lambda )\frac{(\omega _{c}\tau _{0})^{2}-1}{%
		(\omega _{c}\tau _{0})^{2}},  \label{BxxRav} \\
	B_{xy}\approx \omega _{c}\tau _{0}-\frac{R_{T}R_{W}}{\omega _{c}\tau _{0}}%
	\frac{\widetilde{\rho }}{\rho _{0}}-\frac{6R_{D0}^{2}J_{0}^{2}(\lambda )}{%
		\omega _{c}\tau _{0}}.  \label{BxyRav}
\end{gather}%
As can be seen from Eq. (\ref{BxxRav}), for this averaging procedure 
the SO of the diagonal component $%
R_{xx}$ of resistivity tensor vanish at $\omega _{c}\tau _{0}=1$, corresponding to $\pi /\lambda \approx 0.2$ in Fig. \ref{FigRav}, while the
amplitude of the SO of its Hall component $R_{xy}$ does not have zeros. In a strong field at 
$\omega _{c}\tau _{0}>1$ the SO of $R_{xx}$ and $R_{xy}$ have opposite phase, similar to MQO. 
In a weak field at $\omega _{c}\tau _{0}<1$ the SO of $R_{xx}$ and $R_{xy}$ have the same phase, 
but in this limit the amplitude of SO $\sim R_{D0}^2=\exp \left( -2\pi /(\omega _{c}\tau _{0}) \right) \ll 1$ is small.

\subsection{Resistivity MQO with the corrections due to finite interlayer
	electron dispersion}

Contrary to the slow oscillations, the MQO of resistivity tensor in the
first order of Dingle factor do not depend on the averaging procedure and
are given by the second terms in Eqs. (\ref{RtensorXX}) and (\ref{RtensorXY}%
), where $\widetilde{\rho }/\rho _{0}$ is given by Eq. (\ref{TildeRho}).
Combining Eqs. (\ref{R})-(\ref{RtensorXY}) and (\ref{TildeRho}) we obtain
the main MQO term of in-plane resistivity tensor 
\begin{equation}
	R_{xx}^{QO}\approx \frac{-4}{\sigma _{0}}R_{D0}R_{T}R_{W}J_{0}(\lambda )\cos
	\left( \overline{\alpha }\right) ,  \label{RxxMT}
\end{equation}%
\begin{equation}
	R_{xy}^{QO}\approx \frac{2R_{D0}R_{T}R_{W}}{\sigma _{0}\omega _{c}\tau _{0}}%
	J_{0}(\lambda )\cos \left( \overline{\alpha }\right) .  \label{RxyMT}
\end{equation}%
These formulas take only the first main term in Eqs. (\ref{sxxOsc}), (\ref%
{sIxy1}) and (\ref{sIIxy1}), neglecting the corrections $\sim \left( \lambda
/\overline{\alpha }\right) \sim 2t_{z}/\mu $ and $\sim (\gamma +1)/\overline{%
	\alpha }$.

The small correction $\Delta _{\lambda }\hat{\sigma}^{QO}\sim \left( \lambda
/\overline{\alpha }\right) \hat{\sigma}\sim 2t_{z}/\mu $ to conductivity
tensor is obtained from Eqs. (\ref{sxxOsc}) and (\ref{sxyL}): 
\begin{gather}
	\frac{\Delta _{\lambda }\hat{\sigma}^{QO}}{\overline{\sigma }_{xx}}=\left( 
	\begin{array}{cc}
		\Lambda _{xx} & \Lambda _{xy} \\ 
		\Lambda _{yx} & \Lambda _{yy}%
	\end{array}%
	\right) ,\,\text{where}  \label{Ls} \\
	\Lambda _{xx}=\Lambda _{yy}\approx 2\left( \lambda /\overline{\alpha }%
	\right) R_{D0}R_{T}R_{W}J_{1}(\lambda )\sin (\overline{\alpha }),
	\label{Lxx} \\
	\Lambda _{xy}=-\Lambda _{yx}\approx 2\frac{R_{D0}R_{T}R_{W}}{\omega _{c}\tau
		_{0}}\frac{\lambda }{\overline{\alpha }}J_{1}(\lambda )\sin (\overline{%
		\alpha }).  \label{Lxy}
\end{gather}%
We keep this correction for three reasons. First, it is not small when $%
4t_{z}\sim \mu $, i.e. when the $k_{z}$ electron dispersion bandwidth is not
much smaller than the Fermi energy. This case may be relevant to iron-based
and cuprate high-Tc superconductors. The former have several small
Fermi-surface (FS) pockets, warped due to the $k_{z}$ electron dispersion.
The latter also have small FS pockets, as observed by MQO \cite%
{Doiron-Leyraud2007May,Sebastian2015,HelmPRL2009}, coming from the FS
reconstruction due to a charge-density wave \cite{Sebastian2015} or
antiferromagnetic ordering \cite{HelmPRL2009}. Second, the correction $\sim
\lambda /\overline{\alpha }$ gives a measurable phase shift of MQO, because
in contains $\sin (\overline{\alpha })$ instead of $\cos (\overline{\alpha })
$ in the main term. Third, it results to a nonzero MQO amplitude in the beat
nodes. Below we omit the second small correction $\Delta _{\gamma }\hat{%
	\sigma}^{QO}\sim \hat{\sigma}\left( \gamma _{0}+1\right) /\overline{\alpha }$%
, given by Eq. (\ref{sxyQOAdd}), as it is important only in semimetals with
very few occupied LL, when $\hbar \omega _{c}\sim \mu $, and it does not
lead to the phase shift of MQO.

The corrections to resistivity in the first order in $\lambda /\overline{%
	\alpha} =2t_{z}/\mu$ is also in the first order in Dingle factor $R_{D0}$.
Hence, it does not depend on the temperature averaging procedure, similar to
the usual MQO in the first order in $R_{D0}$. The inversion of conductivity
tensor $\hat{\sigma}+\Delta _{\lambda }\hat{\sigma}^{QO}$, similar to that
in the previous subsection, gives the correction to resistivity $\Delta
_{\lambda }\hat{R}$, where the diagonal element%
\begin{equation}
	\Delta _{\lambda }R_{xx}^{QO}\approx\frac{2}{\sigma _{0}}\frac{\lambda }{%
		\overline{ \alpha }}R_{D0}R_{T}R_{W}J_{1}(\lambda )\sin (\overline{\alpha }),
	\label{DLRxx}
\end{equation}%
and the off-diagonal element%
\begin{equation}
	\Delta _{\lambda }R_{xy}^{QO}\approx\frac{-2}{\sigma _{0}}\frac{\lambda }{%
		\overline{ \alpha }}\frac{R_{D0}R_{T}R_{W}}{\omega _{c}\tau _{0}}%
	J_{1}(\lambda )\sin ( \overline{\alpha }).  \label{DLRxy}
\end{equation}

Combining Eqs. (\ref{RxxMT}) and (\ref{RxyMT}) with Eqs. (\ref{DLRxx}) and (%
\ref{DLRxy}) we obtain%
\begin{equation}
	R_{xx}^{QO}\approx \frac{-4R_{D0}R_{T}R_{W}}{\sigma _{0}}\sqrt{1+\left( 
		\frac{\lambda J_{1}(\lambda )}{2\overline{\alpha }J_{0}(\lambda )}\right)
		^{2}}J_{0}(\lambda )\cos \left( \overline{\alpha }+\phi _{xx}\right) ,
	\label{RxxMQO}
\end{equation}%
where the phase shift $\phi _{xx}$ is 
\begin{equation}
	\phi _{xx}=\arctan \left( \frac{\lambda J_{1}(\lambda )}{2\overline{\alpha }%
		J_{0}(\lambda )}\right) ,  \label{PhiXX}
\end{equation}%
and%
\begin{equation}
	R_{xy}^{QO}\approx 2\frac{R_{D0}R_{T}R_{W}}{\omega _{c}\tau _{0}\sigma _{0}}%
	\sqrt{1+\left( \frac{\lambda J_{1}(\lambda )}{\overline{\alpha }%
			J_{0}(\lambda )}\right) ^{2}}J_{0}(\lambda )\cos \left( \overline{\alpha }%
	+\phi _{xy}\right) ,  \label{RxyMQO}
\end{equation}%
where the phase shift $\phi _{xy}$ is 
\begin{equation}
	\phi _{xy}=\arctan \left( \frac{\lambda J_{1}(\lambda )}{\overline{\alpha }%
		J_{0}(\lambda )}\right) .  \label{PhiXY}
\end{equation}
At $ \lambda /\overline{\alpha }= 2t_{z}/\mu \ll 1$ the phase shifts 
$\phi _{xx}<\phi _{xy}\ll 1$ and Eqs. (\ref{RxxMQO})-(\ref{PhiXY}) 
give the same MQO as Eqs. (\ref{R})-(\ref{Rtensor0}). 

Note that the MQO phase shift $\phi _{xy}$ of Hall resistance is almost
twice larger than the MQO phase shift $\phi _{xx}$ of diagonal resistance
components. This can be checked experimentally, if the diagonal $R_{xx}$ and
Hall $R_{xy}$ magnetoresistance oscillations are measured simultaneously.
The calculated correction $\Delta _{\lambda }\hat{R}^{QO}\sim \left(
2t_{z}/\mu\right) \hat{R}$ in resistivity is important even if it is small,
because (i) it gives a measurable phase shift of resistivity MQO given by
Eqs. (\ref{PhiXX}) and (\ref{PhiXY}), and (ii) it gives a nonzero MQO
amplitude even in the beat nodes where $J_{0}(\lambda )=0$.

\begin{figure}
	\centering
	\includegraphics[width=0.5\textwidth]{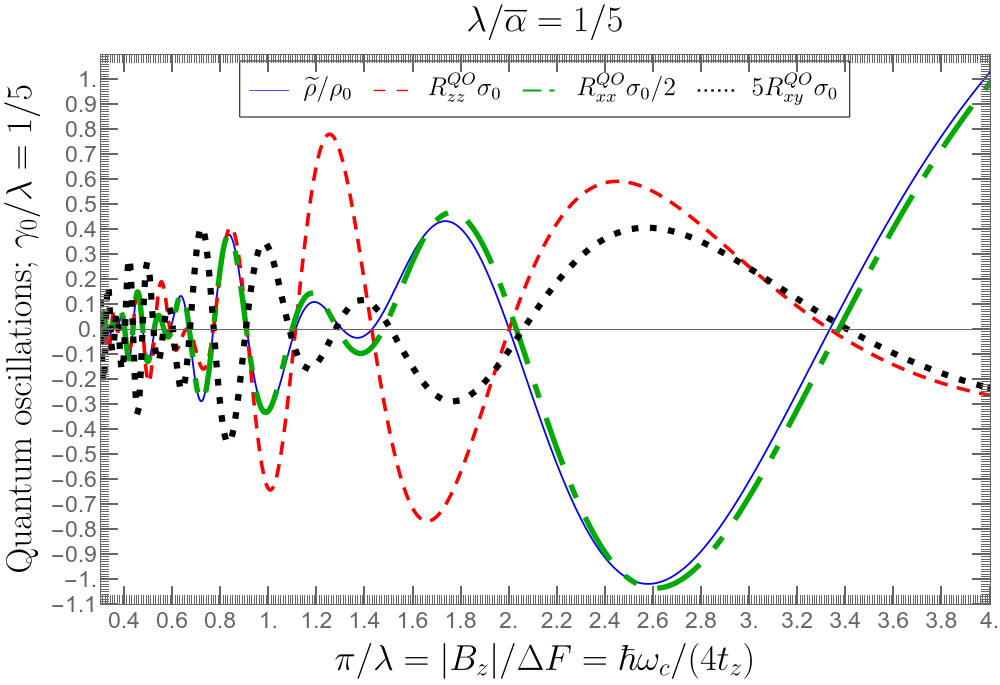} 
\caption{Magnetic quantum oscillations of the density of states $\widetilde{\rho}$ and of resistivity across $R^{QO}_{zz}$ and along the conducting layers, $R^{QO}_{xx}$ and $R^{QO}_{xy}$, given by Eqs. (\ref{TildeRho}), (\ref{RQO}) and (\ref{RxxMQO})-(\ref{PhiXY}) correspondingly, at $\gamma_{0}/\lambda=1/5$ and $\lambda/\overline{\alpha}=1/5$. The MQO of $\widetilde{\rho}$ and $R^{QO}_{xx}$ are in phase, while the MQO of $R^{QO}_{xy}$ are antiphase with them. The MQO of $R^{QO}_{zz}$ have the same phase as $\widetilde{\rho}$ and $R^{QO}_{xx}$ in a weak magnetic field, but at $B_z\sim \Delta F$ the MQO of $R^{QO}_{zz}$ experience a 3D-2D crossover with a phase inversion according to Eq. (\ref{RQO}) and Ref. \cite{Mogilyuk20243d2d}. The MQO amplitude of $R^{QO}_{xx}$ and $R^{QO}_{xy}$ at the beat nodes are not very small, being $\sim\lambda/\overline{\alpha}=1/5$ of their amplitude far from the beat nodes.}
\label{RMQOF6}
\end{figure}

The MQO of diagonal $R^{QO}_{xx}$ and Hall $R^{QO}_{xy}$ resistivity, given by Eqs. (\ref{RxxMQO})-(\ref{PhiXY}), are plotted in Fig. \ref{RMQOF6} in comparison with the MQO of DoS $\widetilde{\rho}$ and of interlayer resistivity $R^{QO}_{zz}$, given by Eqs. (\ref{TildeRho}) and (\ref{RQO}). The parameter $\lambda/\overline{\alpha}=1/5$ in Fig. \ref{RMQOF6} is chosen rather large to illustrate the non-zero amplitude at the beat nodes. The MQO of $\widetilde{\rho}$ and $R^{QO}_{xx}$ are in phase, while the MQO of $R^{QO}_{xy}$ are antiphase with them. The MQO of $R^{QO}_{zz}$ have the same phase as $\widetilde{\rho}$ and $R^{QO}_{xx}$ in a weak magnetic field, but at $B_z\sim \Delta F$ the MQO of $R^{QO}_{zz}$ experience a 3D-2D crossover with a considerable phase shift of beats according to Eq. (\ref{RQO}) and Ref. \cite{Mogilyuk20243d2d}. Therefore, the phase of $R^{QO}_{zz}$ is opposite to the phase of $\widetilde{\rho}$ and $R^{QO}_{xx}$ in wide $B_z$ intervals, e.g. at $B_z > 0.15 \Delta F$  in Fig. \ref{RMQOF6}. At $B_z > \Delta F$ the MQO phase of $R^{QO}_{zz}$ always coincides with that of DoS $\widetilde{\rho}$, but we don't reach this interval in Fig. \ref{RMQOF6}.

\section{Discussions} \label{SecDiscuss}

For the experimental study of the in-plane electronic transport in quasi-2D
metals, especially in a magnetic field, usually, one applies the
van-der-Pauw method \cite{VanderPauw1958,Webster1998measurement} or its
extension -- the Montgomery method \cite{Montgomery1971}. Both these techniques
measure the in-plane resistivity tensor rather than conductivity. Therefore, our 
calculations of the resistivity tensor performed in Sec. \ref{SecR} are necessary 
to describe the experimental data on MQO and SO in quasi-2D metals. This requires 
the calculation of Hall conductivity in addition to its diagonal component 
calculated earlier \cite{Mogilyuk2018Jul}, which is performed in Sec. \ref{SecCond}. 
There is a vast amount of experiments on in-plane magnetoresistance in layered
quasi-2D metals, including various high-Tc superconductors \cite%
{Doiron-Leyraud2007May,ColdeaReview2013}, rare-earth tritellurides \cite%
{Grigoriev2016Jun,Sinchenko2017LinearMag,GrigorievPRB2019}, graphite flakes \cite%
{GraphiteFlackes}, and many other materials. Our analytical results are useful and convenient
to analyze these experimental data.

To calculate the resistivity tensor, which we find by the inversion of
conductivity tensor, $\hat{R} =\hat{\sigma}^{-1}$, one also needs to perform
the averaging over temperature and over macroscopic spatial inhomogeneities. The
macroscopic spatial inhomogeneities result to a spatial variation of the
Fermi energy, which is the difference between the chemical potential and the
bottom of conducting band. The result for the first harmonic of quantum
oscillations does not depend on the order of these two operations: the inversion of
conductivity tensor and its averaging. However,
the difference (slow) oscillations, even if they are stronger than MQO, appear 
in the second order of the Dingle factor, and their amplitude and phase
depend strongly on the order of these two operations.

The temperature averaging physically corresponds to parallel conductivity
channels due to electrons having different energy withing the temperature
smearing of Fermi level. Hence, probably, to calculate the resistivity tensor one
first needs to perform the temperature averaging of conductivity tensor, and
then to invert it. However, for the averaging over macroscopic spatial
inhomogeneity the order may be opposite, if the regions of different Fermi
energy form a series connection of resistances. In real compounds the
regions of various Fermi energy are connected randomly, which is
intermediate between the parallel and series connection of resistances, and one 
can apply various effective medium approximations \cite{Torquato2002} to describe this regime. 
For a 2D chess-board or triangular pattern of two types of resistances $R_{1,\,2}$ 
the effective diagonal resistance component is their geometrical average 
\cite{Marikhin2000}, $R_{eff}=(R_{1}R_{2})^{1/2}$, 
but in a magnetic field this rule may violate because of a large Hall component. 

As shown in Sec. \ref{SubSecRzz}, a similar issue about the 
averaging of conductivity or resistivity emerges for the monotonic growth 
of interlayer magnetoresistance in a strong magnetic field \cite%
{Grigoriev2012Oct}, which is also a non-linear effect in MQO. Averaging
(over MQO period) of interlayer conductivity and resistance gives different
results. Note that the averaging of the experimental data on interlayer resistivity gives slightly better 
agreement between the theory and experiment in quasi-2D organic metal (see Figs. 1 and 2 of 
\cite{Grigoriev2012Oct}), but this observation does not give a general rule of the preferable averaging method.

We suggest that this question about the correct averaging procedure 
of magnetoresistance can be answered experimentally for any particular 
physical system, and our formulas in 
Sec. \ref{SecR} derived for both types of averaging may help to do this. 
The SO of $R_{zz}$ described by Eq. (\ref{RzzSO2}) are smaller than those in Eq. (\ref%
{RzzSO1}) by a factor $\sim 2/\lambda =\hbar\omega_c/(2\pi t_z)$ and 
have a considerably different phase. This phase of SO can be compared with 
the phase of the beats of the MQO in Eqs. (\ref{RxxMQO}) and (\ref{RxyMQO}). 
The phase difference $\Delta \phi_{SO}$ of SO given by Eqs. (\ref{RzzSO1}) 
and (\ref{RzzSO2}), corresponding to $\sigma $-averaging and $R$-averaging,
is rather large: $\Delta \phi_{SO}\sim \pi/2$.
This can be used to experimentally choose the type of averaging, which 
describes better a physical system under study. The comparison of 
the field dependence of the amplitudes and phases of SO of the diagonal $R_{xx}$ and
Hall $R_{xy}$  resistivity components, described by Eqs. (\ref{R})-(\ref{RtensorXY}) 
and (\ref{BxxRav})-(\ref{BxyRav}), also helps to determine which type of averaging 
is more suitable for a particular system.

The combined analysis of
the magnetic quantum and slow (difference) oscillations of the diagonal 
and Hall components of intralayer magnetoresistance tensor and of 
interlayer longitudinal magnetoresistance provides much more useful 
information about the electronic properties of quasi-2D metals than 
the measurements of MQO of a single quantity.  
The analytical formulas for these effects, obtained in Sec. \ref{SecR}, 
are convenient for the comparison with experimental data. 
The results obtained also give several qualitative and quantitative predictions 
on the magnetoresistance oscillations, their relative amplitudes and phases,
which can be tested experimentally. 


\section{Summary} \label{SecSum}

Summing up, in Sec. \ref{SecCond} we calculated the magnetic quantum and slow oscillations 
of intralayer Hall conductivity $\sigma_{xy}$ in quasi-2D metals in a 
quantizing magnetic field, assuming that the oscillating part is still much 
smaller than the monotonic part.
This calculation is based on the Kubo-Streda formula. The
theory takes into account the electron scattering by short-range crystal defects, e.g. by impurities,
and neglects the electron-electron interaction. The latter approximation is
justified in the metallic limit of a large number of filled LLs and finite
interlayer transfer integral $t_z$. Previously, similar calculation in 
quasi-2D metals was performed only for the diagonal components of 
conductivity \cite{Grigoriev2003Apr,Mogilyuk2018Jul}. 
In \ref{SecR} we calculate the intralayer magnetoresistance tensor, 
which allows a direct comparison with experimental data. The averaging 
of magnetoresistance over the period of quantum oscillations due to a finite 
temperature and spatial sample inhomogeneities is nontrivial, as discussed 
in Sec. \ref{SecR} and \ref{SecDiscuss}, which is important both for differential 
slow oscillations and for the monotonic part of magnetoresistance. 

The obtained analytical formulas describe the amplitude and phase of the 
magnetic quantum and slow (difference) oscillations of magnetoresistance tensor as 
a function of magnetic field strength, temperature, disorder, interlayer transfer 
integral, and other parameters. They are convenient for the comparison with experiment 
to extract the electronic parameters of various layered metals. They also predict several 
qualitative and quantitative feature of magnetoresistance oscillations, which can be 
directly compared with experiment. For example, we predict the non-monotonic field 
dependence of the amplitude and phase of slow oscillations in a certain range 
of parameters, the field-dependent phase difference between the oscillations 
of diagonal and Hall magnetoresistance components, the difference between the 
MQO of intralayer and interlayer magnetoresistance, etc.
The developed theory and its results are useful for describing transport
properties in a variety of anisotropic quasi-2D metals, including
high-temperature superconductors, rare-earth polytellurides, layered chalcogenides, 
van-der-Waals crystals, heterostructures, organic metals, etc.

\section{Acknowledgments}

The work is supported by the Russian Science Foundation Grant No. 22-42-09018. PDG also acknowledges State assignment No. FFWR-2024-0015 and NUST "MISIS" grant No. K2-2022-025.

\appendix


\section{Calculation of $\protect\sigma _{xy}^{II}$}\label{App1}

To obtain Eq. (\ref{sIIN}) we integrate the second term in Eq. (\ref{s}%
) by parts

\begin{gather}
\sigma _{\mu \nu }^{II} =\frac{-\left\vert e\right\vert e}{4\pi i}%
\int_{-\infty }^{+\infty }d\varepsilon \,n_{F}\left( \varepsilon \right) 
\nonumber \\
\times Tr\left\langle -2\pi i\frac{d\delta \left( \varepsilon -\hat{H}%
\right) }{d\varepsilon }\left( \hat{r}_{\mu }\hat{v}_{\nu }-\hat{v}_{\mu }%
\hat{r}_{\nu }\right) \right\rangle  \nonumber \\
=\frac{\left\vert e\right\vert e}{2}\int_{-\infty }^{+\infty }d\varepsilon
\,n_{F}\left( \varepsilon \right) Tr\left\langle \frac{d\delta \left(
\varepsilon -\hat{H}\right) }{dB}\frac{\left( \hat{r}_{\mu }\hat{v}_{\nu }-%
\hat{v}_{\mu }\hat{r}_{\nu }\right) }{-d\hat{H}/dB}\right\rangle .
\end{gather}%
For any dispersion with Peierls substitution of the vector potential in the
symmetric gauge $\mathbf{A}=\frac{B_{z}}{2}\left( -y,\,x,\,0\right) $ the
derivative of the Hamiltonian 
\begin{equation}
\frac{d\hat{H}}{dB_{z}}=\frac{e}{2c}\left( \hat{v}_{x}y-\hat{v}_{y}x\right) ,
\end{equation}%
which coincides with $\left( -e/2c\right) \left( \hat{r}_{\mu }\hat{v}_{\nu
}-\hat{v}_{\mu }\hat{r}_{\nu}\right) $ since $\left[ \hat{r}_{x},\,\hat{v}%
_{y}\right] =\left[ \hat{r}_{y},\,\hat{v}_{x}\right] =0$. Using $%
N(\varepsilon )=\int_{-\infty }^{+\infty }d\varepsilon n_{F}\left(
\varepsilon \right) Tr\left\langle \delta \left( \varepsilon -\hat{H}\right)
\right\rangle $ one obtains Eq. (\ref{sIIN}). This derivation does not
depend on electron dispersion and on the number of energy bands.

\section{Calculation of $\protect\sigma _{xy}^{I}$}\label{App2}

In the self-consistent Born approximation (SCBA) or even in the non-crossing
approximation, i.e. neglecting only the diagrams with the intersection of
impurity lines, the self-energy part $\Sigma ^{R}(\varepsilon )$ depends
only on electron energy $\varepsilon $ \cite{Grigoriev2011Jun}, and the
electron Green's function does not depend on $k_{x}$:%
\begin{equation}
G_{n}^{R}(k_{x},\,k_{z},\,\varepsilon)=G_{n}^{A}(k_{x},\,k_{z},\,\varepsilon
)^*=\frac{1}{\varepsilon -\epsilon _{n}(k_z)-\Sigma ^{R}(\varepsilon)},
\label{G}
\end{equation}
where $\epsilon _{n}(k_z)=\hbar \omega _{c}(n+1/2)-2t_z\cos(k_zd)$ according to Eq. (\ref{ES}). The
imaginary part of the Green's function, entering the expression (\ref{sI})  for conductivity, is%
\begin{gather}
\text{Im}G_{n}^{R}(k_{x},\,k_{z},\,\varepsilon )\equiv \text{Im}%
G_{n}^{R}(k_{z},\,\varepsilon )=-\text{Im}G_{n}^{A}(k_{z},\,\varepsilon ) 
\nonumber \\
=(G_n^R(k_{z},\,\varepsilon ) -G_n^A(k_{z},\,\varepsilon ) )/(2i)  \nonumber
\\
=\frac{\text{Im}\Sigma ^{R}(\varepsilon )}{\left[ \varepsilon -\epsilon
_{n}(k_z)-\text{Re}\Sigma ^{R}(\varepsilon )\right] ^{2}+\left[ \text{Im}%
\Sigma ^{R}(\varepsilon )\right] ^{2}}.  \label{ImG}
\end{gather}
The matrix elements $\left\langle n^{\prime },\,k_{x},\,k_{z}|\hat{v}%
_{i}|k_{z},\,k_{x},\,n\right\rangle $ of electron velocity $%
v_{i}=p_{i}/m_{\ast }$, entering to the electric current $%
j_{i }=ev_{i }=e\, \partial \epsilon \left( \mathbf{p}\right) /\partial
p_{i}$, in the basis of the Landau-gauge quantum numbers $%
\left\{ k_{x},\,k_{z},\,n\right\} $ of an electron in magnetic field are 
\cite{Streda1975Aug}%
\begin{gather}
\left\langle n^{\prime },\,k_{x},\,k_{z}|\hat{v}_{x}|k_{z},\,k_{x},\,n\right%
\rangle  \nonumber \\
=\frac{-\hbar }{\sqrt{2}m_{\ast }l_{H}}\left( \sqrt{n^{\prime }+1}\delta
_{n,\,n^{\prime }+1}+\sqrt{n^{\prime }}\delta _{n,\,n^{\prime }-1}\right)
\label{vx} \\
=\frac{-\hbar }{\sqrt{2}m_{\ast }l_{H}}\left( \sqrt{n}\delta _{n,\,n^{\prime
}+1}+\sqrt{n+1}\delta _{n,\,n^{\prime }-1}\right) ,
\end{gather}%
\begin{gather}
\left\langle n^{\prime },\,k_{x},\,k_{z}|\hat{v}_{y}|k_{z},\,k_{x},\,n\right%
\rangle  \nonumber \\
=\frac{i\hbar }{\sqrt{2}m_{\ast }l_{H}}\left( \sqrt{n^{\prime }+1}\delta
_{n,\,n^{\prime }+1}-\sqrt{n^{\prime }}\delta _{n,\,n^{\prime }-1}\right)
\label{vy} \\
=\frac{i\hbar }{\sqrt{2}m_{\ast }l_{H}}\left( \sqrt{n}\delta _{n,\,n^{\prime
}+1}-\sqrt{n+1}\delta _{n,\,n^{\prime }-1}\right) ,
\end{gather}%
or%
\begin{gather}
\left\langle n,\,k_{x},\,k_{z}|\hat{v}_{x}|k_{z},\,k_{x},\,n^{\prime
}\right\rangle  \nonumber \\
=\frac{-\hbar }{\sqrt{2}m_{\ast }l_{H}}\left( \sqrt{n+1}\delta _{n^{\prime
}-1,\,n}+\sqrt{n}\delta _{n^{\prime }+1,\,n}\right),
\end{gather}%
\begin{gather}
\left\langle n,\,k_{x},\,k_{z}|\hat{v}_{y}|k_{z},\,k_{x},\,n^{\prime
}\right\rangle  \nonumber \\
=\frac{i\hbar }{\sqrt{2}m_{\ast }l_{H}}(\sqrt{n+1}\delta _{n^{\prime }-1,\,n}-%
\sqrt{n}\delta _{n,\,n^{\prime }+1}),
\end{gather}%
where $l_{H}=\sqrt{\hbar c/(eB)}=\sqrt{\hbar /(m_{\ast }\omega _{c})}$ is
the magnetic length. Eqs. (\ref{vx}) and (\ref{vy}) can be checked by
a direct calculation. 

The nonzero matrix elements of electron velocity correspond to the change of LL number only by $\pm 1$:
\begin{equation}
	\left\langle n|\hat{v}_{x}|\,n-1\right\rangle =\frac{-\hbar \sqrt{n}}{\sqrt{2%
		} m_{\ast }l_{H}};~\left\langle n|\hat{v}_{x}|\,n+1\right\rangle =\frac{%
		-\hbar \sqrt{ n+1}}{\sqrt{2}m_{\ast }l_{H}},  \label{vx1}
\end{equation}
and for the $\hat{v}_{y}$ component%
\begin{equation}
	\left\langle n|\hat{v}_{y}|\,\,n-1\right\rangle =\frac{-i\hbar \sqrt{n}}{%
		\sqrt{2} m_{\ast }l_{H}};~\left\langle n|\hat{v}_{y}|\,n+1\right\rangle =%
	\frac{i\hbar \sqrt{ n+1}}{\sqrt{2}m_{\ast }l_{H}}.  \label{vy1}
\end{equation}
From Eqs. (\ref{vx1}) and (\ref{vy1}) one easily
obtains the product of the matrix elements of electron velocity:
\begin{equation}
\left\langle n|\hat{v}_{x}|\,n-1\right\rangle \left\langle n-1|\hat{v}%
_{y}|\,n\right\rangle =\frac{-i\hbar ^{2}n}{2m_{\ast }^{2}l_{H}^{2}},
\label{v2xm}
\end{equation}%
\begin{equation}
\left\langle n|\hat{v}_{x}|\,n+1\right\rangle \left\langle n+1|\hat{v}%
_{y}|\,n\right\rangle =\frac{i\hbar ^{2}\left( n+1\right) }{2m_{\ast
}^{2}l_{H}^{2}}.  \label{v2xp}
\end{equation}%
Substituting $\hat{\jmath}_{\mu }=e\hat{v}_{\mu }$ into the Eq. (\ref{sI}%
) leads to 
\begin{equation}
\sigma _{xy}^{I}\left( \varepsilon \right) =\frac{\hbar e^{2}}{4\pi }
Tr\left\langle \hat{v}_{x}\left( G^R-G^A\right) \hat{v}_{y}G^R-\hat{v}_{x}G^A%
\hat{v}_{y}\left( G^R-G^A\right) \right\rangle .  \label{sIxy}
\end{equation}
The first term in the Eq. (\ref{sIxy}) gives 
\begin{gather}
\left\langle n|\hat{v}_{x}\left(G^R-G^A\right) \hat{v}_{y}G^R |n\right\rangle
\nonumber \\
=\left\langle n|\hat{v}_{x}|\,n-1\right\rangle \left\langle n-1|\hat{v}%
_{y}|\,n\right\rangle G_{n}^{R}\left( G_{n-1}^{R}-G_{n-1}^{A}\right) 
\nonumber \\
+\left\langle n|\hat{v}_{x}|\,n+1\right\rangle \left\langle n+1|\hat{v}%
_{y}|\,n\right\rangle G_{n}^{R}\left( G_{n+1}^{R}-G_{n+1}^{A}\right) 
\nonumber \\
=\frac{-i\hbar ^{2}}{2m_{\ast }^{2}l_{H}^{2}}n\left(
G_{n-1}^{R}-G_{n-1}^{A}\right) G_{n}^{R}  \nonumber \\
+\frac{i\hbar ^{2}}{2m_{\ast }^{2}l_{H}^{2}}\left( n+1\right) \left(
G_{n+1}^{R}-G_{n+1}^{A}\right) G_{n}^{R}.  \label{Mn1}
\end{gather}%
The second term in the Eq. (\ref{sIxy}) gives 
\begin{gather}
\left\langle n|\hat{v}_{x}G^A\hat{v}_{y}\left( G^R-G^A\right) |n\right\rangle
\nonumber   \\
=\left\langle n|\hat{v}_{x}|\,n-1\right\rangle \left\langle n-1|\hat{v}%
_{y}|\,n\right\rangle G_{n-1}^{A}\left( G_{n}^{R}-G_{n}^{A}\right)  \nonumber
\\
+\left\langle n|\hat{v}_{x}|\,n+1\right\rangle \left\langle n+1|\hat{v}%
_{y}|\,n\right\rangle G_{n+1}^{A}\left( G_{n}^{R}-G_{n}^{A}\right)  \nonumber
\\
=\frac{-i\hbar ^{2}}{2m_{\ast }^{2}l_{H}^{2}}n\left(
G_{n}^{R}-G_{n}^{A}\right) G_{n-1}^{A}  \nonumber \\
+\frac{i\hbar ^{2}}{2m_{\ast }^{2}l_{H}^{2}}\left( n+1\right) \left(
G_{n}^{R}-G_{n}^{A}\right) G_{n+1}^{A}.  \label{Mn2}
\end{gather}%
With the help of Eqs. (\ref{Mn1}) and (\ref{Mn2}) one can rewrite the
Eq. (\ref{sI}) as 
\begin{gather}
\sigma _{xy}^{I}=\frac{\hbar e^{2}}{4\pi }Tr\left\langle \hat{v}_{x}\left(
G^R-G^A\right) \hat{v}_{y}G^R-\hat{v}_{x}G^A\hat{v}_{y}\left( G^R-G^A\right)
\right\rangle  \nonumber \\
=\frac{ie^{2}\hbar ^{3}}{8\pi m_{\ast }^{2}l_{H}^{2}}\sum_{n,\,k_z}\left[
-n\left( G_{n-1}^{R}-G_{n-1}^{A}\right) G_{n}^{R}\right.  \nonumber \\
+\left( n+1\right) \left( G_{n+1}^{R}-G_{n+1}^{A}\right) G_{n}^{R}+n\left(
G_{n}^{R}-G_{n}^{A}\right) G_{n-1}^{A}  \nonumber \\
\left. -\left( n+1\right) \left( G_{n}^{R}-G_{n}^{A}\right) G_{n+1}^{A} 
\right]  \label{Mn2t}
\end{gather}
In this formula the Green's functions $G_{n}^{R,\,A}$ depend both on the LL
number $n$ and on the band index $k_z$, because the dispersion $\epsilon
_{n}(k_z)$ depends on $k_z$. We now redefine the summation index $%
n\rightarrow n+1$ in the first and third terms in the square brackets of Eq. (\ref{Mn2t}), which
gives 
\begin{gather}
\sigma _{xy}^{I} =\frac{ie^{2}\hbar ^{3}}{8\pi m_{\ast }^{2}l_{H}^{2}}
\sum_{n,\,k_z}\left[ -\left( n+1\right) \left( G_{n}^{R}-G_{n}^{A}\right)
G_{n+1}^{R}\right.  \nonumber \\
+\left( n+1\right) \left( G_{n+1}^{R}-G_{n+1}^{A}\right) G_{n}^{R}  \nonumber
\\
+\left( n+1\right) \left( G_{n+1}^{R}-G_{n+1}^{A}\right) G_{n}^{A}  \nonumber
\\
\left. -\left( n+1\right) \left( G_{n}^{R}-G_{n}^{A}\right) G_{n+1}^{A} 
\right].  \label{sgm}
\end{gather}%
Substituting Eqs. (\ref{G}), (\ref{ImG}) into (\ref{sgm}) we get 
\begin{gather}
\sigma _{xy}^{I}=\frac{-e^{2}\hbar ^{3}}{2\pi m_{\ast }^{2}l_{H}^{2}}
\sum_{n,\,k_z }\left( n+1\right) \times \\
\left[ \frac{-\text{Im}\Sigma ^{R}(\varepsilon )}{\left[ \varepsilon
-\epsilon _{n+1}(k_z)-\text{Re}\Sigma ^{R}(\varepsilon )\right] ^{2}+\left[ 
\text{ Im}\Sigma ^{R}(\varepsilon )\right] ^{2}}\right.  \nonumber \\
\times \frac{\varepsilon -\epsilon _{n}(k_z)-\text{Re}\Sigma
^{R}(\varepsilon ) }{\left[ \varepsilon -\epsilon _{n}(k_z)-\text{Re}\Sigma
^{R}(\varepsilon )\right] ^{2}+\left[ \text{Im}\Sigma ^{R}(\varepsilon )%
\right] ^{2}}  \nonumber \\
-\frac{-\text{Im}\Sigma ^{R}(\varepsilon )}{\left[ \varepsilon -\epsilon
_{n}(k_z)-\text{Re}\Sigma ^{R}(\varepsilon )\right] ^{2}+\left[ \text{Im}%
\Sigma ^{R}(\varepsilon )\right] ^{2}}  \nonumber \\
\times \left. \frac{\varepsilon -\epsilon _{n+1}(k_z)-\text{Re}\Sigma
^{R}(\varepsilon )}{\left[ \varepsilon -\epsilon _{n+1}(k_z)-\text{Re}\Sigma
^{R}(\varepsilon )\right] ^{2}+\left[ \text{Im}\Sigma ^{R}(\varepsilon ) %
\right] ^{2}}\right].  \label{sgm-60}
\end{gather}%
The first and the second terms in Eq. (\ref{sgm-60}) have the same denominator, and we simplify it to%
\begin{gather}
\sigma _{xy}^{I} =\frac{e^{2}\hbar ^{3}}{2\pi m_{\ast }^{2}l_{H}^{2}}
\sum_{n,\,k_z }\frac{\left( n+1\right) \text{Im}\Sigma ^{R}(\varepsilon )}{ %
\left[ \varepsilon -\epsilon _{n+1}(k_z)-\text{Re}\Sigma ^{R}(\varepsilon )%
\right] ^{2}+\left[ \text{Im}\Sigma ^{R}(\varepsilon )\right] ^{2}} 
\nonumber \\
\times \frac{\epsilon _{n+1}(k_z)-\epsilon _{n}(k_z)}{\left[ \varepsilon
-\epsilon _{n}(k_z)-\text{Re}\Sigma ^{R}(\varepsilon )\right] ^{2}+\left[ 
\text{Im}\Sigma ^{R}(\varepsilon )\right] ^{2}}.  \label{sxyIc2}
\end{gather}
Using the expression for $\sigma _{xx}$ in Eq. (24) of Ref. \cite%
{Streda1975Aug}, 
\begin{gather}
\sigma _{xx}=\frac{e^{2}\hbar ^{3}}{\pi m_{\ast }^{2}l_{H}^{2}}
\sum_{n,\,k_z }\frac{\left( n+1\right) \text{Im}\Sigma ^{R}(\varepsilon )}{ %
\left[ \varepsilon -\epsilon _{n+1}(k_z)-\text{Re}\Sigma ^{R}(\varepsilon )%
\right] ^{2}+\left[ \text{Im}\Sigma ^{R}(\varepsilon )\right] ^{2}} 
\nonumber \\
\times \frac{\text{Im}\Sigma ^{R}(\varepsilon )}{\left[ \varepsilon
-\epsilon _{n}(k_z)-\text{Re}\Sigma ^{R}(\varepsilon )\right] ^{2}+\left[ 
\text{Im}\Sigma ^{R}(\varepsilon )\right] ^{2}},
\end{gather}%
and noting that $\left( \epsilon _{n+1}(k_z)-\epsilon _{n}(k_z)\right)
/\left\vert \text{Im}\Sigma ^{R}(\varepsilon )\right\vert =2\omega _{c}\tau$
and Im$\Sigma ^{R}(\varepsilon )=-\left\vert \text{Im}\Sigma
^{R}(\varepsilon )\right\vert $, we reduce Eq. (\ref{sxyIc2}) to Eq. (\ref%
{sIN}): $\sigma _{xy}^{I}=-\omega _{c}\tau\sigma _{xx}$.

\section{Harmonic expansion of the electron density of states and the total electron number}\label{App3}

The electron DoS per each spin component is given by the sum over Landau levels: 
\begin{gather}
	\rho _{e}(\varepsilon ) =-\frac{\text{Im}G^R(\varepsilon )}{\pi} =\sum_{n=0}^{+\infty }\sum_{k_{z}}\text{Im}\frac{%
		N_{LL}/\pi }{\varepsilon -\epsilon \left( n,\,k_{z}\right) +i0}  \notag
	\label{rhon1} \\
	=\sum_{n=0}^{+\infty }\frac{N_{LL}/(\pi d)}{\sqrt{4t_{z}^{2}-\left(
			\varepsilon -\hbar \omega _{c}\left(n+1/2\right) \right) ^{2}}},
	\label{rhon}
\end{gather}%
where $N_{LL}\equiv B|e|/(2\pi \hbar c)$ is the LL degeneracy per one spin
component in 2D, coming from the summation over $k_{y}$. When the expression inside the square root becomes negative, the total expression should always be taken equal to zero, because the DoS $\rho_e$ is a real quantity. To find the MQO one
usually transforms sum over LLs in the DoS given by Eq. (\ref{rhon}) to the
sum over harmonics using the Poisson summation formula:%
\begin{equation}
	\sum_{n=n_{0}}^{+\infty }f(n)=\sum_{k=-\infty }^{+\infty }\int_{a}^{+\infty
	}e^{2\pi ikn}f(n)\,dn\,,  \label{Poisson}
\end{equation}%
where $a\in (n_{0}-1,\,n_{0})$. Taking $a=0$, below the lowest $n=0$ LL, one
obtains%
\begin{equation}
	\rho _{e}(\varepsilon )=\frac{\rho _{0}}{\pi }\sum_{k=-\infty }^{+\infty
	}\int_{0}^{+\infty }\frac{dn\,\exp \left( 2\pi ik\left( n-1/2\right) \right) 
	}{\sqrt{\left( \frac{2t_{z}}{\hbar \omega _{c}}\right) ^{2}-\left( \frac{%
				\varepsilon }{\hbar \omega _{c}}-\,n\right) ^{2}}},  \label{rhoInt}
\end{equation}%
where $\rho _{0}=N_{LL}/(\hbar \omega _{c}d)$ is the average DoS at the
Fermi level. To proceed further one assumes the Fermi energy $\mu \gg
t_{z},\,\hbar \omega_{c}$. Since only the LLs with energy $\left\vert \hbar
\omega _{c}(n+1/2)-\mu \right\vert <2t_{z}$ and $n\gg 1$ contribute to the
DoS at the Fermi level $\rho (\mu )$, the latter is not sensitive to the
lower integration limit over $n$. Then to find $\rho (\mu )$ analytically,
one extends the integration over $n$ in Eq. (\ref{rhoInt}) from $\left(
0,\,+\infty \right) $ to $\left(-\infty ,\,+\infty \right) $. As a result one
replaces $\rho _{e}(\varepsilon )$, which is nonzero only at $\varepsilon
>\hbar \omega _{c}/2-2t_{z}$, by $\rho (\varepsilon )$ given by the
well-known formula \cite{Champel2001Jan} 
\begin{gather}
	\rho (\varepsilon )=\sum_{n=-\infty }^{+\infty }\frac{N_{LL}/(\pi d)}{\sqrt{%
			4t_{z}^{2}-\left( \varepsilon -\hbar \omega _{c}\left(n+1/2\right)
			\right) ^{2}}}  \label{rhons} \\
	=\frac{\rho _{0}}{\pi }\sum_{k=-\infty }^{+\infty }\int_{-\infty }^{+\infty }%
	\frac{dn\,\exp \left( 2\pi ik\left( n-1/2\right) \right) }{\sqrt{\left( 
			\frac{2t_{z}}{\hbar \omega _{c}}\right) ^{2}-\left( \frac{\varepsilon }{%
				\hbar \omega _{c}}-\,n\right) ^{2}}}  \notag \\
	=\rho _{0}\sum_{k=-\infty }^{+\infty }\left( -1\right) ^{k}\exp \left( 
	\frac{2\pi ik\varepsilon }{\hbar \omega _{c}}\right) J_{0}\left( \frac{4\pi
		kt}{\hbar \omega _{c}}\right) .  \label{Rhohar}
\end{gather}%
Keeping only zeroth and first harmonics in this expression one obtains Eq. (%
\ref{rho}). Contrary to $\rho _{e}(\varepsilon )$, $\rho (\varepsilon )$
given by Eqs. (\ref{Rhohar}) or (\ref{rho}) is nonzero at\ $\varepsilon
<\hbar \omega _{c}/2-2t_{z}$.

The above extension of the lower integration limit and of the DoS $\rho
_{e}(\varepsilon )\rightarrow \rho (\varepsilon )$ to negative energy $%
\varepsilon $ below the bottom of the conducting band needs some care when
the integrals over DoS $\rho (\varepsilon )$ are involved, as in Eq. (\ref%
{Ntot0}) for the total electron density, or in the calculation of the
thermodynamic potential and magnetization \cite%
{Champel2001Jan,Grigoriev2001Jun,Mogilyuk2022}. Evidently, the result in Eq.
(\ref{Ntot0}) depends on the lower integration limit over $\varepsilon $,
contrary to $a\in (n_{0}-1,\,n_{0})$ in Eqs. (\ref{Poisson}), (\ref{rhoInt}).
The correct starting expression for the total electron density is%
\begin{equation}
	N=\int_{-\infty }^{+\infty }\rho _{e}(\varepsilon )n_{F}\left( \varepsilon
	\right) d\varepsilon \approx \int_{-\infty }^{\mu }\rho _{e}(\varepsilon
	)d\varepsilon   \label{N}
\end{equation}%
with $\rho _{e}(\varepsilon )$ given by Eq. (\ref{rhon}). The last approximate 
equality in Eq. (\ref{N}) holds at low and temperature $T$ as compared
to LL separation $\hbar \omega _{c}$. It also holds at arbitrary $T$ when
the chemical potential is just on the maxima or minima of the DoS $\rho
(\varepsilon )$. However, the last approximate equality in Eq. (\ref{N}) may
slightly violate at $T\sim \hbar \omega _{c}$ when the chemical potential
deviates from the extrema of the DoS $\rho (\varepsilon )$, so that the
latter contains an odd term as a function of $\varepsilon -\mu $. 

The replacement of $\rho _{e}(\varepsilon )$ in Eq. (\ref{N}) by $\rho
(\varepsilon )$ from Eqs. (\ref{Rhohar}) or (\ref{rho}) gives $N=+\infty $.
More importantly, the magnetic field derivative in Eq. (\ref{sIIder}) then
oscillates as a function of the lower integration limit in Eqs. (\ref{Ntot0}%
) or (\ref{N}) with the period of MQO without a definite value. This problem
can be solved by the observation that 
\begin{equation}
	N\approx \int_{-\infty }^{\mu }\rho _{e}(\varepsilon )d\varepsilon
	=\int_{0}^{\mu }\rho (\varepsilon )d\varepsilon .  \label{Nc}
\end{equation}%
The difference 
\begin{equation}
	\delta N=\int_{-\infty }^{\mu }\rho _{e}(\varepsilon )d\varepsilon
	-\int_{0}^{\mu }\rho (\varepsilon )d\varepsilon   \label{dN0}
\end{equation}
after the substitution of Eqs. (\ref{rhon}) and (\ref{rhons}) becomes   
\begin{gather}
	\delta N =\int_{-\infty }^{\mu }\sum_{n=0}^{+\infty }\frac{N_{LL}/(\pi d)}{%
		\sqrt{4t_{z}^{2}-\left( \varepsilon -\hbar \omega _{c}\left(n+1/2\right)
			\right) ^{2}}}d\varepsilon   \label{dN0r} \\
	-\int_{0}^{\mu }\sum_{n=-\infty }^{+\infty }\frac{N_{LL}/(\pi d)}{\sqrt{%
			4t_{z}^{2}-\left( \varepsilon -\hbar \omega _{c}\left(n+1/2\right)
			\right) ^{2}}}d\varepsilon .  \notag
\end{gather}%
Both integrals in Eq. (\ref{dN0r}) contain the same terms%
\begin{equation*}
	\int_{0}^{\mu }\sum_{n=0}^{+\infty }\frac{N_{LL}/(\pi d)}{\sqrt{%
			4t_{z}^{2}-\left( \varepsilon -\hbar \omega _{c}\left(n+1/2\right)
			\right) ^{2}}}d\varepsilon ,
\end{equation*}%
which cancel each other. Then Eq. (\ref{dN0r}) rewrites as%
\begin{gather}
	\delta N=\int_{-\infty }^{0}\sum_{n=0}^{+\infty }\frac{N_{LL}/(\pi d)}{\sqrt{%
			4t_{z}^{2}-\left( \varepsilon -\hbar \omega _{c}\left( n+1/2\right)
			\right) ^{2}}}d\varepsilon   \label{dN1} \\
	-\int_{0}^{\mu }\sum_{n=-\infty }^{-1}\frac{N_{LL}/(\pi d)}{\sqrt{%
			4t_{z}^{2}-\left( \varepsilon -\hbar \omega _{c}\left(n+1/2\right)
			\right) ^{2}}}d\varepsilon .  \notag  \label{dN1r}
\end{gather}%
In quasi-2D metals each LL contributes the electronic DoS in the energy
interval $\left(-2t_{z},\,2t_{z}\right) $ of $k_{z}$-dispersion bandwidth $%
4t_{z}$. Therefore, each sum over $n$ in Eq. (\ref{dN1}) contains only $n_{t}
$ nonzero terms, where $n_{t}$ is the smallest integer greater than or equal
to $2t_{z}/\hbar \omega _{c}-1/2$. Moreover, the sum in the second line of
Eq. (\ref{dN1}) is nonzero only at $\varepsilon <2t_{z}-\hbar \omega
_{c}/2\ll \mu $, and the upper integration limit in the second line of Eq. (%
\ref{dN1}) can be extended from $\mu $ to $+\infty $. Then the variable
change $\varepsilon \rightarrow -\varepsilon $ and $n\rightarrow -n-1$
identically transforms the second integral in Eq. (\ref{dN1}) to the first
one: 
\begin{gather*}
	\int_{0}^{+\infty }\sum_{n=-\infty }^{-1}\frac{N_{LL}/(\pi d)}{\sqrt{%
			4t_{z}^{2}-\left( \varepsilon -\hbar \omega _{c}\left(n+1/2\right)
			\right) ^{2}}}d\varepsilon  \\
	\rightarrow\int_{-\infty }^{0}\sum_{n=0}^{+\infty }\frac{N_{LL}/(\pi d)}{%
		\sqrt{4t_{z}^{2}-\left( -\varepsilon -\hbar \omega _{c}\left(-n-1/2\right) \right) ^{2}}}d\varepsilon  \\
	=\int_{-\infty }^{0}\sum_{n=0}^{+\infty }\frac{N_{LL}/(\pi d)}{\sqrt{%
			4t_{z}^{2}-\left( \varepsilon -\hbar \omega _{c}\left(n+1/2\right)
			\right) ^{2}}}d\varepsilon .
\end{gather*}%
Hence, two terms in  Eq. (\ref{dN1}) exactly cancel each other, giving 
$\delta N=0$\ and proving Eq. (\ref{Nc}). This substantiates Eq. (\ref{Ntot0}) and shows that the low integration limit $\varepsilon =0$ in  Eq. (\ref{Ntot0}) is the only correct.

\bibliographystyle{apsrev4-2}
\bibliography{Papers,Book}

\end{document}